\documentclass[]{pasj01}
\draft

\Received{$\langle$reception date$\rangle$}
\Accepted{$\langle$acception date$\rangle$}
\Published{$\langle$publication date$\rangle$}
\usepackage{natbib}
\usepackage{ulem}
\usepackage{comment}
\usepackage{txfonts}
\usepackage{color} 
\usepackage{lscape}
\DeclareUnicodeCharacter{2212}{-}

\begin{document} 

\title{ 
Intermediate Luminosity Type Iax SN 2019muj With Narrow Absorption Lines: 
Long-Lasting Radiation Associated With a Possible Bound Remnant Predicted 
by the Weak Deflagration Model}
\author{Miho Kawabata\altaffilmark{1},
  Keiichi Maeda\altaffilmark{1},
  Masayuki Yamanaka\altaffilmark{2},
  Tatsuya Nakaoka\altaffilmark{3,4},
  Koji S. Kawabata\altaffilmark{3,4,5},
  Kentaro Aoki\altaffilmark{6},
  G. C. Anupama\altaffilmark{7},
  Umut Burgaz\altaffilmark{1,8}, 
  Anirban Dutta\altaffilmark{7},
  Keisuke Isogai\altaffilmark{2},
  Masaru Kino\altaffilmark{2},
  Naoto Kojiguchi\altaffilmark{1},
  Iida Kota\altaffilmark{9},  
  Brajesh Kumar\altaffilmark{7,10},
  Daisuke Kuroda\altaffilmark{2},
  Hiroyuki Maehara\altaffilmark{11}, 
  Kazuya Matsubayashi\altaffilmark{2},
  Kumiko Morihana\altaffilmark{12},
  Katsuhiro L. Murata\altaffilmark{9},
  Tomohito Ohshima\altaffilmark{13},
  Masaaki Otsuka\altaffilmark{2},
  Devendra K. Sahu\altaffilmark{7},  
  Avinash Singh\altaffilmark{7},
  Koji Sugitani\altaffilmark{14},
  Jun Takahashi\altaffilmark{13},
and  Kengo Takagi\altaffilmark{3}
}%
\altaffiltext{1}{Department of Astronomy, Kyoto University, Kitashirakawa-Oiwakecho, Sakyo-ku, Kyoto 606-8502, Japan}
\altaffiltext{2}{Okayama Observatory, Kyoto University, 3037-5 Honjo, Kamogata-cho, Asakuchi, Okayama 719-0232, Japan}
\altaffiltext{3}{Hiroshima Astrophysical Science Center, Hiroshima University, Higashi-Hiroshima, Hiroshima 739-8526, Japan}
\altaffiltext{4}{Department of Physical Science, Hiroshima University, Kagamiyama 1-3-1, Higashi-Hiroshima 739-8526, Japan}
\altaffiltext{5}{Core Research for Energetic Universe (CORE-U), Hiroshima University, Kagamiyama, Higashi-Hiroshima, Hiroshima 739-8526, Japan}
\altaffiltext{6}{Subaru Telescope, National Astronomical Observatory of Japan, 650 North A’ohoku Place,
Hilo, HI 96720, USA}
\altaffiltext{7}{Indian Institute of Astrophysics, Koramangala 2nd Block, Bengaluru 560034, India}
\altaffiltext{8}{Department of Astronomy and Space Sciences, Ege University, 35100 Izmir, Turkey}
\altaffiltext{9}{Department of Physics, Tokyo Institute of Technology, 2-12-1 Ookayama, Meguro-ku, Tokyo 152-8551, Japan}
\altaffiltext{10}{Aryabhatta Research Institute of Observational Sciences, Manora Peak,
Nainital - 263 001, India}
\altaffiltext{11}{Subaru Telescope Okayama Branch Office, National Astronomical Observatory of Japan, National Institutes of Natural Sciences, 3037-5 Honjo, Kamogata, Asakuchi, Okayama 719-0232, Japan
}
\altaffiltext{12}{Graduate School of Science, Nagoya University, Furo-cho, Chikusa-ku, Nagoya 464-8602, Japan}
\altaffiltext{13}{Nishi-Harima Astronomical Observatory, Center for Astronomy, University of Hyogo, 407-2 Nishigaichi, Sayo-cho, Sayo, Hyogo 679-5313, Japan}
\altaffiltext{14}{Graduate School of Science, Nagoya City University, Mizuho-ku, Nagoya, Aichi 467-8601, Japan}
\email{kawabata@kusastro.kyoto-u.ac.jp}


\maketitle

\begin{abstract}
We present comprehensive spectroscopic and photometric analyses of the 
intermediate luminosity Type Iax supernova (SN Iax) 2019muj
based on multi-band datasets observed through the framework of the OISTER
target-of-opportunity program.
SN 2019muj exhibits almost identical characteristics with the subluminous 
SNe Iax 2008ha and 2010ae in terms of the observed spectral features and 
the light curve evolution at the early phase, except for the peak luminosity.
The long-term observations unveil the flattening light curves at the
late time as seen in a luminous SN Iax 2014dt.
This can be explained by the existence of an inner dense
and optically-thick component possibly associated with a bound white
dwarf remnant left behind the explosion.
We demonstrate that the weak deflagration model with a wide range of 
the explosion parameters can reproduce the late-phase light curves of 
other SNe Iax.
Therefore, we conclude that a common explosion mechanism operates for 
different subclass SNe Iax.
\end{abstract}

\section{Introduction}\label{sec:intro}
It has been widely accepted that type Ia supernovae 
(SNe Ia) arise from the thermonuclear runaway of a massive C+O white 
dwarf (WD) in a binary system. 
There is however ongoing discussion on the natures of the progenitor 
and explosion mechanism; the progenitor WD may either a 
Chandrasekhar-limiting mass WD or a sub-Chandrasekhar mass WD, and 
the thermonuclear runaway may be ignited either near the center or 
the surface of the WD \citep[see, e.g., ][for a review]{Maeda2016}.
For SNe Ia, there is a well-established correlation between the peak 
luminosity and the light-curve decline rate, which is known as the 
luminosity-width relation (\citealt{Phillips93}). 
This relation allows SNe Ia to be used as precise standardized candles 
to measure the cosmic-scale distances to remote galaxies and thus the 
cosmological parameters (\citealt{Riess98}; \citealt{Perlmutter99}).

A class of peculiar SNe Ia has been discovered since the early 2000s. 
Their peak absolute magnitudes are significantly dimmer than those 
expected from the correlation. 
Their properties should be specified to avoid the sample contamination 
for cosmological studies.
These outliers have been called SN 2002cx-like SNe (\citealt{Li03}) 
or SNe Iax (\citealt{Foley13}).
SNe Iax commonly show lower luminosities (M$_{R} \sim$ $-14$ -- $-19$ mag), 
lower expansion velocities (2,000 --  8,000 km s$^{−1}$), and lower 
explosion energies (10$^{49}$ -- 10$^{51}$ erg) than normal SNe Ia (e.g., 
\citealt{Foley09}; \citealt{Foley13}; \citealt{Stritzinger15}).
As an extreme example, the maximum magnitude of the faintest SN Iax 
2008ha is only $−13.74 \pm 0.15$ mag in the $B$ band (\citealt{Foley09}), 
which is $\sim$ 4 mag fainter than that of the prototypical SN Iax 
SN 2002cx. 

Many researchers have discussed plausible models that can reproduce 
the wide range of the observed properties of SNe Iax (e.g., \citealt{Hoeflich95}; \citealt{Hoeflich96}; 
\citealt{Nomoto76}; \citealt{Jha06}; \citealt{Sahu08}; \citealt{Moriya10}). 
One of such models is a weak deﬂagration of a 
Chandrasekhar-mass carbon-oxygen WD (e.g., \citealt{Jordan12}, 
\citealt{Kromer13}, \citealt{Fink14}), in which a considerable part 
of the WD could not gain sufﬁcient kinetic energy to exceed the 
binding energy, leaving a bound WD remnant after the explosion. 
\cite{Kromer15} suggested, in the context of the weak deflagration 
model, that the faintest SNe Iax can be explained 
by a hybrid oxygen-neon WD with a carbon-oxygen core.
While the weak deflagration is a popular scenario (e.g., \citealt{Jha17}), 
it has not yet been established. 
One important test we suggest in the present work (see also 
\citealt{Kawabata18}) is the late-time light curve behavior, 
which was indeed not a scope of the original light curve models by 
\cite{Fink14}. 
To consider the possible observational signature associated with 
the bound remnant with a range of properties, the investigation of the 
light curves to the late-phase is highly important.

To date, the number of SNe Iax for which the evolution of the late phases 
has been well-observed remains small.
The decline rates in their late phases may potentially add further 
diversity; they may not even correlate with the properties in the early 
phases (\citealt{Stritzinger15}; \citealt{Yamanaka15}; \citealt{Kawabata18}).

SN 2019muj (ASASSN-19tr) was discovered at 17.4 mag on 2019 Aug. 7.4 
UT by All Sky Automated Survey for SuperNovae (ASAS-SN, 
\citealt{Shappee14}) in the nearby galaxy VV 525 (\citealt{Brimacombe19}).
The spectrum was obtained on Aug. 7.8 UT by the Las Cumbres 
Observatory Global SN project and this SN was classified as an SN Iax 
at about a week before the maximum light (\citealt{Hiramatsu19}). 
\citet{Barna21} reported the early phase observations 
($\sim$ 50 days).
They analyzed the spectra by comparing them with the synthetic spectra 
obtained from radiative transfer calculations.
In this paper, we report the extended multi-band 
observation of SN Iax 2019muj, especially focusing on the analysis of its 
late-time light curve behavior; being discovered in the very nearby galaxy, 
the SN gives us an opportunity to obtain optical and
near-infrared data up to 200 days after the explosion.
We describe the observations and data reduction in \S\ref{sec:obs}.
We present the results of the observations and compare the properties 
of SN 2019muj with those of other SNe Iax in \S\ref{sec:results}. 
SN 2019muj is classified as an intermediate SN Iax, and shows the slow 
evolution of the light curve at 100--200 days.
In \S\ref{sec:discussion}, we discuss whether or not the weak deflagration  
can explain the observational properties of SN 2019muj.
We also apply our analytical methods to other SNe Iax and demonstrate 
that our modified weak deflagration scenario can account for their 
light curves.
A summary of this work is provided in \S\ref{sec:conclusions}.

\section{Observations and Data Reduction}\label{sec:obs}

We performed spectral observations of SN 2019muj using Hiroshima
One-shot Wide-field Polarimeter (HOWPol; \citealt{Kawabata08}) 
mounted on the 1.5 m Kanata telescope of Hiroshima University, 
the Kyoto Okayama Optical Low-dispersion Spectrograph with an 
integral field unit (KOOLS-IFU; \citealt{Yoshida05}, 
\citealt{Matsubayashi19}) on the 3.8m Seimei telescope of Kyoto 
University (\citealt{Kurita20}), 
Hanle Faint Object Spectrograph (HFOSC) mounted 
on the 2 m Himalayan Chandra Telescope (HCT) of the Indian 
Astronomical Observatory, 
and the Faint Object Camera and Spectrograph (FOCAS; 
\citealt{Kashikawa02}) 
installed at the 8.2 m Subaru telescope, NAOJ.
Multi-band imaging observations were conducted as a 
Target-of-Opportunity (ToO) program in the framework of the 
Optical and Infrared Synergetic Telescopes for Education and 
Research (OISTER).
All the magnitudes given in this paper are in the Vega system.

\subsection{Photometry}\label{sec:photo}
We performed $BVRI$-band imaging observations using HOWPol and 
$VRI$-band imaging observations using the Hiroshima Optical and 
Near-InfraRed camera (HONIR; \citealt{Akitaya14}) installed at the 
Kanata telescope. 
We also obtained $UBVRI$-band images data with the 2 m HCT.
Additionally, $VR$-band imaging observation was performed using FOCAS 
installed at the Subaru telescope. 

\begin{figure}
 \begin{center}
  \includegraphics[width=10cm]{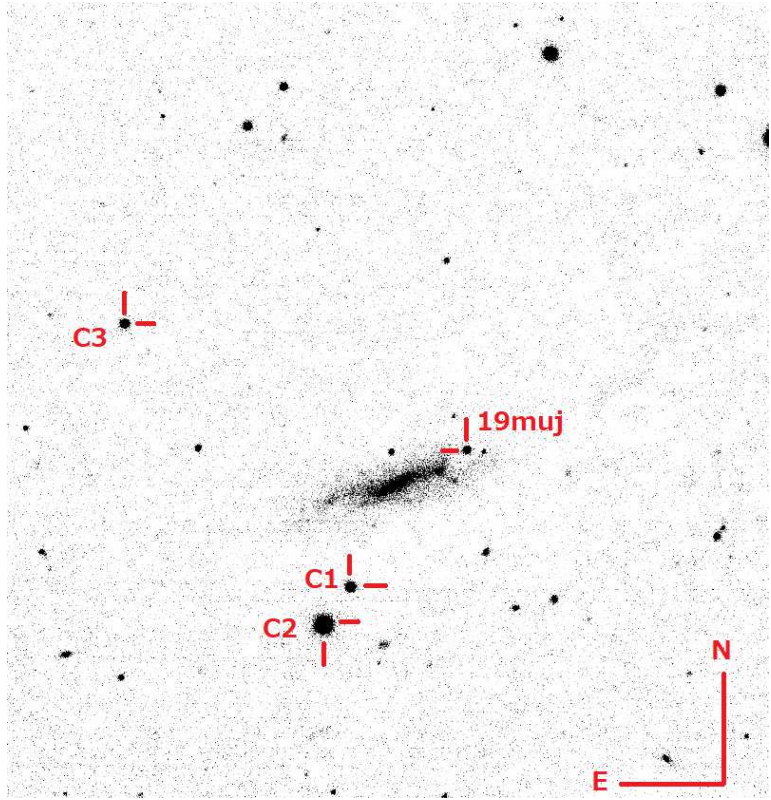} 
 \end{center}
\caption{$R$-band image of SN 2019muj and the comparison stars taken 
with the Kanata telescope / HOWPol on MJD 58725.76 (2019 Aug. 30).}
\label{fig:cmp_star}
\end{figure}

We reduced the imaging data in a standard manner for the CCD 
photometry. 
The journal of the optical photometry is listed in Table \ref{tb:opt_log}. 
We adopted the Point-Spread-Function (PSF) fitting photometry method 
using DAOPHOT package in {\it IRAF}\footnote{{\it IRAF} 
is distributed by the National Optical Astronomy Observatory, which 
is operated by the Association of Universities for Research in 
Astronomy (AURA) under a cooperative agreement with the National 
Science Foundation.}. 
We skipped the S-correction, since it is negligible for 
the purposes of the present study (\citealt{Stritzinger02}).
For the magnitude calibration, we adopted relative photometry using the 
comparison stars taken in the same frames (Figure \ref{fig:cmp_star}). 
The magnitudes of the comparison stars in the $BVRI$ bands  
were calibrated with the stars in the UGC 11860 field (\citealt{Singh18}) 
observed on a photometric night, as shown in Table \ref{tb:cmp_star}.
The secondary standard stars in this field were calibrated 
using the Landolt photometric standards (\citealt{Landolt92}).
First-order color term correction was applied in the photometry. 
The amount of the correction for the color term is, 
for example, $\sim$ 0.03 mag in the case of HOWPol in $B$ band 
at the maximum light. 
In the longer wavelength bands, the amount of the color-term correction 
is smaller. 
While the amount of this correction varies depending on the instruments 
and the color of the object, it is generally negligible for the purpose 
of the present study.

In deriving the SN magnitudes, we did not perform the galaxy template-image 
subtraction method. 
We have checked the contamination from the host galaxy using the 
pre-discovery images obtained by  Pan-STARRS\footnote{https://outerspace.stsci.edu/display/PANSTARRS/Pan-STARRS1+data+archive+home+page}. 
Within an aperture having a diameter of $\sim$ 2" (as a typical seeing 
for the Kanata observations) centered on the SN position, the background 
magnitudes are estimated to be $m_{g}$ = 21.58 mag, $m_{r}$ = 21.97 mag, 
or $m_{i}$ = 21.57 mag. 
The contamination is thus negligible throughout the observation period.
Indeed, the light curves keep declining toward the late phase, which supports 
that the contamination by the possible background emission is negligible. 
Therefore, it would not affect the conclusions in this paper.

\begin{table}
  \tbl{Magnitudes of comparison stars of SN 2019muj.}{%
  \begin{tabular}{cccccccc}
      \hline
      ID &  $B$  &  $V$  &  $R$  & $I$   & $J$\footnotemark[$*$]   & $H$\footnotemark[$*$]   & $K_{s}$\footnotemark[$*$]\\ 
      \hline
      C1 & 17.258 $\pm$ 0.030 & 16.552 $\pm$ 0.023 & 16.076 $\pm$ 0.017 & 15.671 $\pm$ 0.031 & 15.206 $\pm$ 0.043 & 14.880 $\pm$ 0.069 & 14.592 $\pm$ 0.093 \\
      C2 & 15.426 $\pm$ 0.027 & 14.638 $\pm$ 0.019 & 14.134 $\pm$ 0.017 & 13.742 $\pm$ 0.030 & 13.294 $\pm$ 0.023 & 12.919 $\pm$ 0.021 & 12.861 $\pm$ 0.029 \\
      C3 & 17.786 $\pm$ 0.037 & 16.974 $\pm$ 0.021 & 16.465 $\pm$ 0.018 & 16.060 $\pm$ 0.030 & 16.369 $\pm$ 0.105 & 15.923 $\pm$ 0.143 & 15.474 $\pm$ 0.195 \\
      \hline
    \end{tabular}}\label{tb:cmp_star}
    \begin{tabnote}
    \footnotemark[$*$] The magnitudes in the NIR bands are from the 2MASS catalog (\citealt{Persson98}).  \\ 
\end{tabnote}
\end{table}

\begin{longtable}{ccccccccc}
\caption{Log of optical observations of SN 2019muj.}\label{tb:opt_log}
\hline
  Date  & MJD  &  Phase\footnotemark[$\dag$]  & $U$   & $B$   &  $V$   & $R$   &  $I$  & Telescope \\
          &      &  (day)                       & (mag) & (mag) & (mag)  & (mag) & (mag) & (Instrument) \\
\endfirsthead
\hline
  Date  & MJD  &  Phase\footnotemark[$\dag$]  & $U$   & $B$   &  $V$   & $R$   &  $I$  & Telescope \\
        &      &  (day)                       & (mag) & (mag) & (mag)  & (mag) & (mag) & (Instrument) \\
\hline
\endhead
  \hline
\endfoot
  \hline
\endlastfoot
  \hline
2019-08-08 & 58703.8 & $-3.9$ &           --        &          --        & 17.122 $\pm$ 0.098 & 17.020 $\pm$ 0.111 & 16.932 $\pm$ 0.109 & Kanata (HONIR) \\
2019-08-09 & 58704.7 & $-3.0$ &           --        & 16.947 $\pm$ 0.075 & 16.911 $\pm$ 0.054 & 16.630 $\pm$ 0.049 & 16.713 $\pm$ 0.053 & Kanata (HOWPol) \\
2019-08-09 & 58704.8 & $-2.9$ &           --        &          --        & 16.586 $\pm$ 0.066 & 16.409 $\pm$ 0.085 & 16.421 $\pm$ 0.093 & Kanata (HONIR) \\
2019-08-11 & 58706.7 & $-1.0$ &           --        &          --        &          --        & 16.477 $\pm$ 0.023 & 16.467 $\pm$ 0.034 & Kanata (HOWPol) \\
2019-08-11 & 58706.8 & $-0.9$ &           --        &          --        & 16.471 $\pm$ 0.079 & 16.332 $\pm$ 0.097 & 16.352 $\pm$ 0.099 & Kanata (HONIR) \\
2019-08-12 & 58707.7 &   0.0  &           --        &          --        & 16.173 $\pm$ 0.063 & 16.032 $\pm$ 0.059 & 15.976 $\pm$ 0.060 & Kanata (HONIR) \\
2019-08-13 & 58708.7 &   1.1  &           --        &          --        & 16.507 $\pm$ 0.068 & 16.316 $\pm$ 0.058 & 16.370 $\pm$ 0.030 & Kanata (HONIR) \\
2019-08-13 & 58708.8 &   1.1  &           --        & 16.677 $\pm$ 0.041 &         --         & 16.247 $\pm$ 0.029 & 16.200 $\pm$ 0.069 & Kanata (HOWPol) \\
2019-08-30 & 58725.7 &  18.0  &           --        &          --        & 17.593 $\pm$ 0.036 & 17.158 $\pm$ 0.020 & 16.813 $\pm$ 0.045 & Kanata (HONIR) \\
2019-08-30 & 58725.8 &  18.1  &           --        & 19.350 $\pm$ 0.076 & 17.667 $\pm$ 0.026 & 17.094 $\pm$ 0.055 & 16.803 $\pm$ 0.036 & Kanata (HOWPol) \\
2019-09-04 & 58730.8 &  23.1  &           --        &          --        & 17.849 $\pm$ 0.074 & 17.217 $\pm$ 0.109 &          --        & Kanata (HONIR) \\
2019-09-04 & 58730.8 &  23.1  &           --        & 19.396 $\pm$ 0.030 & 17.890 $\pm$ 0.022 & 17.361 $\pm$ 0.021 & 17.005 $\pm$ 0.032 & Kanata (HOWPol) \\
2019-09-06 & 58732.7 &  25.1  &           --        & 19.628 $\pm$ 0.030 & 17.999 $\pm$ 0.024 & 17.442 $\pm$ 0.024 & 17.047 $\pm$ 0.031 & Kanata (HOWPol) \\
2019-09-06 & 58732.8 &  25.1  &           --        &          --        & 17.976 $\pm$ 0.092 & 17.475 $\pm$ 0.095 & 17.165 $\pm$ 0.067 & Kanata (HONIR) \\
2019-09-14 & 58740.7 &  33.1  &           --        &          --        & 18.344 $\pm$ 0.122 & 17.970 $\pm$ 0.146 & 17.487 $\pm$ 0.156 & Kanata (HONIR) \\
2019-09-14 & 58740.8 &  33.1  &           --        &          --        &          --        &           --       & 17.215 $\pm$ 0.030 & Kanata (HOWPol) \\
2019-09-15 & 58741.8 &  34.1  &           --        &          --        & 18.374 $\pm$ 0.186 & 17.803 $\pm$ 0.180 & 17.388 $\pm$ 0.118 & Kanata (HONIR) \\
2019-09-15 & 58741.8 &  34.1  &           --        &          --        & 18.305 $\pm$ 0.051 & 17.750 $\pm$ 0.082 & 17.427 $\pm$ 0.042 & Kanata (HOWPol) \\
2019-09-16 & 58742.7 &  35.0  &           --        &          --        & 18.215 $\pm$ 0.205 & 17.831 $\pm$ 0.128 & 17.425 $\pm$ 0.081 & Kanata (HONIR) \\
2019-09-16 & 58742.7 &  35.0  &           --        &          --        & 18.467 $\pm$ 0.081 & 17.954 $\pm$ 0.034 & 17.427 $\pm$ 0.030 & Kanata (HOWPol) \\
2019-09-23 & 58749.8 &   42.1 &           --        & 19.696 $\pm$ 0.012 & 18.461 $\pm$ 0.011 & 17.994 $\pm$ 0.005 & 17.479 $\pm$ 0.012 & HCT (HFOSC) \\
2019-09-24 & 58750.6 &  42.9  &           --        & 19.710 $\pm$ 0.030 & 18.499 $\pm$ 0.059 & 18.018 $\pm$ 0.032 & 17.566 $\pm$ 0.033 & Kanata (HOWPol) \\
2019-09-25 & 58751.6 &  43.9  &           --        & 19.848 $\pm$ 0.102 & 18.568 $\pm$ 0.045 & 18.021 $\pm$ 0.042 & 17.555 $\pm$ 0.066 & Kanata (HOWPol) \\
2019-09-26 & 58752.7 &  45.0  &           --        & 20.264 $\pm$ 0.246 & 18.626 $\pm$ 0.028 & 17.957 $\pm$ 0.066 &           --       & Kanata (HOWPol) \\
2019-10-01 & 58758.9 &   51.2 &           --        & 19.799 $\pm$ 0.023 & 18.605 $\pm$ 0.008 & 18.185 $\pm$ 0.010 & 17.730 $\pm$ 0.011 & HCT (HFOSC) \\ 
2019-10-03 & 58759.6 &  51.9  &           --        & 19.954 $\pm$ 0.030 & 18.662 $\pm$ 0.041 & 18.192 $\pm$ 0.021 & 17.736 $\pm$ 0.031 & Kanata (HOWPol) \\
2019-10-04 & 58760.7 &  53.0  &           --        & 20.105 $\pm$ 0.036 & 18.724 $\pm$ 0.029 & 18.224 $\pm$ 0.021 & 17.762 $\pm$ 0.031 & Kanata (HOWPol) \\
2019-10-06 & 58762.6 &  55.0  &           --        &          --        & 18.565 $\pm$ 0.160 & 18.252 $\pm$ 0.106 & 17.816 $\pm$ 0.063 & Kanata (HONIR) \\
2019-10-06 & 58762.7 &  55.1  &           --        & 20.271 $\pm$ 0.224 & 18.803 $\pm$ 0.048 & 18.295 $\pm$ 0.037 & 17.794 $\pm$ 0.049 & Kanata (HOWPol) \\
2019-10-08 & 58764.6 &  56.9  &           --        &          --        & 18.644 $\pm$ 0.113 & 18.035 $\pm$ 0.170 & 17.638 $\pm$ 0.208 & Kanata (HONIR) \\
2019-10-08 & 58764.7 &  57.0  &           --        & 20.114 $\pm$ 0.268 & 18.789 $\pm$ 0.039 & 18.297 $\pm$ 0.028 & 17.820 $\pm$ 0.039 & Kanata (HOWPol) \\
2019-10-11 & 58767.9 &   60.2 &           --        &          --        & 18.818 $\pm$ 0.015 & 18.313 $\pm$ 0.010 & 17.899 $\pm$ 0.012 & HCT (HFOSC) \\
2019-10-16 & 58772.7 &  65.0  &           --        & 19.727 $\pm$ 0.030 & 19.183 $\pm$ 0.122 & 18.530 $\pm$ 0.060 & 17.956 $\pm$ 0.056 & Kanata (HOWPol) \\
2019-10-22 & 58778.6 &  71.0  &           --        &          --        & 18.801 $\pm$ 0.076 & 18.299 $\pm$ 0.090 & 17.707 $\pm$ 0.393 & Kanata (HONIR) \\
2019-10-22 & 58778.6 &  70.9  &           --        & 20.255 $\pm$ 0.153 & 19.006 $\pm$ 0.029 & 18.547 $\pm$ 0.026 & 18.030 $\pm$ 0.057 & Kanata (HOWPol) \\
2019-10-29 & 58785.6 &  77.9  &           --        & 20.254 $\pm$ 0.038 & 19.196 $\pm$ 0.075 & 18.639 $\pm$ 0.020 & 18.072 $\pm$ 0.035 & Kanata (HOWPol) \\
2019-10-30 & 58786.9 &   79.2 &           --        &          --        & 19.085 $\pm$ 0.023 & 18.662 $\pm$ 0.023 & 18.198 $\pm$ 0.028 & HCT (HFOSC) \\
2019-11-06 & 58789.7 &  82.0  &           --        & 20.222 $\pm$ 0.030 & 19.549 $\pm$ 0.020 & 18.715 $\pm$ 0.045 & 18.148 $\pm$ 0.040 & Kanata (HOWPol) \\
2019-11-05 & 58792.9 &   85.2 &           --        & 20.100 $\pm$ 0.040 & 19.182 $\pm$ 0.024 & 18.681 $\pm$ 0.023 & 18.217 $\pm$ 0.032 & HCT (HFOSC) \\
2019-11-11 & 58794.6 &  87.0  &           --        & 19.957 $\pm$ 0.030 & 19.610 $\pm$ 0.271 & 18.928 $\pm$ 0.160 & 18.304 $\pm$ 0.090 & Kanata (HOWPol) \\
2019-11-25 & 58808.6 & 100.9  &           --        &          --        & 19.536 $\pm$ 0.058 & 18.944 $\pm$ 0.050 & 18.313 $\pm$ 0.039 & Kanata (HOWPol) \\
2019-12-08 & 58825.5 & 117.9  &           --        &          --        & 19.862 $\pm$ 0.033 & 19.315 $\pm$ 0.035 &        --         & Subaru (FOCAS) \\
2019-12-13 & 58830.6 & 122.9  &           --        &          --        & 19.974 $\pm$ 0.119 & 19.129 $\pm$ 0.048 & 18.381 $\pm$ 0.049 & Kanata (HOWPol) \\
2019-12-28 & 58845.5 & 137.8  &           --        &          --        &          --        & 19.123 $\pm$ 0.042 & 18.470 $\pm$ 0.036 & Kanata (HOWPol) \\
2020-02-28 & 58907.6 & 199.9  &           --        &          --        & 20.348 $\pm$ 0.056 & 19.634 $\pm$ 0.064 &          --        & HCT (HFOSC) \\
\end{longtable}

We also performed $JHK_{\rm s}$-band imaging observations with HONIR 
attached to the Kanata telescope, with the Nishi-harima 
Infrared Camera (NIC) installed at the Cassegrain focus of the 2.0 m 
Nayuta telescope at the NishiHarima Astronomical Observatory, and with 
the Near-infrared simultaneous three-band camera (SIRIUS; 
\citealt{Nagayama03}) installed at the 1.4m IRSF telescope at the 
South African Astronomical Observatory.
We adopted the sky background subtraction using a template sky image 
obtained by the dithering observation. 
We performed the PSF fitting photometry in the same way as used for the 
reduction of the optical data and calibrated the magnitude using the 
comparison stars in the 2MASS catalog (\citealt{Persson98}). 
We list the journal of the NIR photometry in Table \ref{tb:nir_log}.

\begin{table}
  \tbl{Log of NIR observations of SN 2019muj.}{%
  \begin{tabular}{ccccccc}
      \hline
        Date  & MJD  &  Phase\footnotemark[$\ddag$]  & $J$   & $H$   & $K_{\rm s}$ & Telescope     \\
              &      &  (day)                      & (mag) & (mag) & (mag)       & (Instrument)   \\
      \hline
2019-08-08 & 58703.7 & $-3.9$ & 17.228 $\pm$ 0.069 &          --        &          --        & Kanata (HONIR) \\
2019-08-09 & 58704.8 & $-2.9$ & 16.927 $\pm$ 0.064 & 17.059 $\pm$ 0.093 &          --        & Kanata (HONIR) \\
2019-08-10 & 58705.0 & $-2.6$ & 16.920 $\pm$ 0.045 & 16.944 $\pm$ 0.071 & 17.023 $\pm$ 0.108 & IRSF (SIRIUS) \\
2019-08-11 & 58706.0 & $-1.7$ & 16.765 $\pm$ 0.047 & 16.698 $\pm$ 0.075 &         --         & IRSF (SIRIUS) \\
2019-08-11 & 58706.8 & $-0.9$ & 16.661 $\pm$ 0.053 & 16.923 $\pm$ 0.109 &          --        & Kanata (HONIR) \\
2019-08-12 & 58707.7 &   0.0  & 16.560 $\pm$ 0.060 & 16.753 $\pm$ 0.113 &          --        & Kanata (HONIR) \\
2019-08-13 & 58708.7 &   1.0  & 16.809 $\pm$ 0.065 &          --        &          --        & Kanata (HONIR) \\
2019-08-12 & 58707.8 &   1.1  & 16.443 $\pm$ 0.043 & 17.484 $\pm$ 0.073 &        --          & Nayuta (NIC)  \\
2019-08-14 & 58709.0 &   1.3  & 16.715 $\pm$ 0.047 & 16.616 $\pm$ 0.075 & 16.570 $\pm$ 0.114 & IRSF (SIRIUS) \\
2019-08-16 & 58711.0 &   3.3  & 16.795 $\pm$ 0.045 & 16.567 $\pm$ 0.072 & 16.501 $\pm$ 0.113 & IRSF (SIRIUS) \\
2019-08-17 & 58712.0 &   4.3  &          --        & 16.432 $\pm$ 0.083 & 16.526 $\pm$ 0.102 & IRSF (SIRIUS) \\
2019-08-22 & 58717.0 &   9.3  & 16.981 $\pm$ 0.045 & 16.359 $\pm$ 0.071 & 16.562 $\pm$ 0.103 & IRSF (SIRIUS) \\
2019-08-24 & 58719.0 &  11.3  & 17.034 $\pm$ 0.044 & 16.398 $\pm$ 0.070 & 16.648 $\pm$ 0.098 & IRSF (SIRIUS) \\
2019-08-27 & 58722.0 &  14.3  & 17.072 $\pm$ 0.044 & 16.503 $\pm$ 0.070 & 17.084 $\pm$ 0.102 & IRSF (SIRIUS) \\
2019-08-30 & 58725.7 &  18.0  &          --        & 16.405 $\pm$ 0.116 &          --        & Kanata (HONIR) \\
2019-09-02 & 58728.0 &  20.3  & 17.329 $\pm$ 0.044 & 16.770 $\pm$ 0.071 & 17.010 $\pm$ 0.106 & IRSF (SIRIUS) \\
2019-09-02 & 58728.9 &  21.2  & 17.377 $\pm$ 0.045 & 16.830 $\pm$ 0.071 & 17.141 $\pm$ 0.116 & IRSF (SIRIUS) \\
2019-09-03 & 58729.9 &  22.2  & 17.401 $\pm$ 0.046 & 16.832 $\pm$ 0.073 &        --          & IRSF (SIRIUS) \\
2019-09-04 & 58730.8 &  23.1  & 17.540 $\pm$ 0.113 &          --        &          --        & Kanata (HONIR) \\
2019-09-14 & 58740.7 &  33.0  &          --        & 17.029 $\pm$ 0.142 &          --        & Kanata (HONIR) \\
2019-09-15 & 58741.7 &  34.1  &          --        & 17.167 $\pm$ 0.108 &          --        & Kanata (HONIR) \\
2019-09-16 & 58742.7 &  35.0  &          --        & 17.043 $\pm$ 0.166 &          --        & Kanata (HONIR) \\
2019-10-08 & 58764.6 &  56.9  & 17.838 $\pm$ 0.174 &          --        &          --        & Kanata (HONIR) \\
2019-10-22 & 58778.6 &  70.9  & 18.681 $\pm$ 0.244 & 18.254 $\pm$ 0.196 &          --        & Kanata (HONIR) \\
\hline
\end{tabular}}\label{tb:nir_log}
\begin{tabnote}
\footnotemark[$\ddag$] Relative to the epoch of $B$-band maximum (MJD 58707.69).  \\ 
\end{tabnote}
\end{table}

Additionally, we downloaded the imaging data obtained by $Swift$ 
Ultraviolet/Optical Telescope (UVOT) from the $Swift$ Data 
Archive\footnote{http://www.swift.ac.uk/swift\_portal/}.
In the UV data of $Swift$, we adopted the absolute photometry using 
the zero points reported by \cite{Breeveld11}.
We performed PSF fitting photometry using {\it IRAF} for these data.
We list the journal of the UV photometry in Table \ref{tb:uv_log}.

\begin{table}
  \tbl{Log of $Swift$/UVOT observations of SN 2019muj.}{%
  \begin{tabular}{cccccccc}
      \hline
        Date  & MJD  &  Phase\footnotemark[$\S$]  & $V$   & $B$   & $U$   & $uvw1$ & $uvw2$ \\
              &      &  (day)  & (mag) & (mag) & (mag) & (mag)  & (mag)    \\
      \hline
2019-08-07 & 58702.8 & -4.8 & 17.183 $\pm$ 0.139 & 17.406 $\pm$ 0.078 & 16.212 $\pm$ 0.053 & 16.750 $\pm$ 0.068 & 17.919 $\pm$ 0.093 \\
2019-08-08 & 58703.8 & -3.8 & 16.752 $\pm$ 0.097 & 16.952 $\pm$ 0.060 & 15.885 $\pm$ 0.038 & 16.846 $\pm$ 0.051 & 17.814 $\pm$ 0.089 \\
2019-08-09 & 58704.6 & -3.1 & 16.390 $\pm$ 0.073 & 16.787 $\pm$ 0.048 & 15.752 $\pm$ 0.037 & 16.991 $\pm$ 0.051 &        --          \\
2019-08-10 & 58705.4 & -2.3 & 16.472 $\pm$ 0.078 & 16.641 $\pm$ 0.046 & 15.915 $\pm$ 0.041 & 17.032 $\pm$ 0.053 & 17.906 $\pm$ 0.074 \\
2019-08-11 & 58706.8 & -0.9 & 16.602 $\pm$ 0.090 & 16.537 $\pm$ 0.045 & 15.860 $\pm$ 0.043 & 17.323 $\pm$ 0.065 & 18.196 $\pm$ 0.103 \\
2019-08-15 & 58710.7 & 3.0 & 16.470 $\pm$ 0.085 & 16.743 $\pm$ 0.052 & 16.463 $\pm$ 0.058 & 17.914 $\pm$ 0.086 & 18.662 $\pm$ 0.137 \\
2019-08-17 & 58712.9 & 5.2 & 16.483 $\pm$ 0.082 & 16.978 $\pm$ 0.060 & 17.079 $\pm$ 0.080 & 18.417 $\pm$ 0.108 & 19.030 $\pm$ 0.164 \\
2019-08-25 & 58720.0 & 12.4 & 17.213 $\pm$ 0.151 & 18.374 $\pm$ 0.176 & 18.663 $\pm$ 0.278 & 19.736 $\pm$ 0.231 & 19.131 $\pm$ 0.219 \\
2019-08-29 & 58724.3 & 16.6 & 17.062 $\pm$ 0.160 & 18.407 $\pm$ 0.161 & 18.902 $\pm$ 0.273 &        --        & 20.414 $\pm$ 0.518 \\
2019-09-03 & 58729.6 & 21.9 & 17.319 $\pm$ 0.138 & 18.965 $\pm$ 0.246 &        --          &        --        &        --        \\
2019-09-04 & 58730.0 & 22.3 & 17.552 $\pm$ 0.162 & 18.799 $\pm$ 0.208 &        --          &        --        &        --        \\
      \hline
    \end{tabular}}\label{tb:uv_log}
    \begin{tabnote}
    \footnotemark[$\S$] Relative to the epoch of $B$-band maximum (MJD 58707.69).  \\ 
\end{tabnote}
\end{table}

\subsection{Spectroscopy}\label{sec:spec}
For the spectra taken with HOWPol, the wavelength coverage is 
$4500$--$9200$ \AA \, and the wavelength 
resolution is $R = \lambda/\Delta\lambda \simeq 400$ at 6000 \AA. 
For wavelength calibration, we used sky emission lines. 
To remove cosmic ray events, we used the {\it L. A. Cosmic} 
pipeline (\citealt{Dokkum01}; \citealt{Dokkum12}). 
The flux of SN 2019muj was calibrated using the data of 
spectrophotometric standard stars that have taken on the same night. 

The spectra with KOOLS-IFU installed on Seimei telescope were taken 
through the optical fibers. 
We used the VPH-blue grism.
The wavelength coverage is $4000$--$8900$ \AA \ and the wavelength 
resolution is $R = \lambda/\Delta\lambda \sim 500$. 
The data reduction was performed using Hydra package in {\it IRAF} 
(\citealt{Barden94}; \citealt{Barden95}) and a reduction software 
specifically developed for KOOLS-IFU data.
A sky frame was separately taken, which is then subtracted from 
the object frame.
For the wavelength calibration, we used the arc lamp (Hg and Ne) data.

For the spectrum obtained with FOCAS, the wavelength coverage is 
$3700$--$10000$ \AA \ and the wavelength resolution is 
$R = \lambda/\Delta\lambda \sim 650$ at 6000 \AA. 
The data reduction was performed in the same way as that with 
HOWPol, except that we used the arc lamp (Th-Ar) data 
and skylines for the wavelength calibration.
The journal of spectroscopy is listed in Table \ref{tb:spec_log}.

\begin{table}
  \tbl{Log of the spectroscopic observations of SN 2019muj.}{%
  \begin{tabular}{cccccc}
      \hline
        Date  & MJD  &  Phase\footnotemark[$\|$]  & Coverage  & Resolution  & Telescope  \\
             &      &  (day)  & (\AA)     & (\AA)       & (Instrument)\\
      \hline
2019-08-08 & 58703.7 & $-4.0$ & $4000$--$8900$  & 500 & Seimei (KOOLS)  \\
2019-08-10 & 58705.7 & $-2.0$ & $4000$--$8900$  & 500 & Seimei (KOOLS)  \\
2019-08-11 & 58706.7 & $-1.0$ & $4000$--$8900$  & 500 & Seimei (KOOLS)  \\
2019-08-11 & 58706.8 & $-0.9$ & $4500$--$9200$  & 400 & Kanata (HOWPol) \\
2019-08-16 & 58711.7 &   4.1  & $4000$--$8900$  & 500 & Seimei (KOOLS)  \\
2019-08-17 & 58712.7 &   5.0  & $4000$--$8900$  & 500 & Seimei (KOOLS)  \\
2019-08-21 & 58716.7 &   9.0  & $4000$--$8900$  & 500 & Seimei (KOOLS)  \\
2019-09-05 & 58731.7 &  24.0  & $4000$--$8900$  & 500 & Seimei (KOOLS)  \\
2019-09-23 & 58749.8 &  42.1  & $3500$--$9000$  & 300 & HCT  (HFOSC) \\
2019-09-24 & 58750.7 &  43.0  & $4000$--$8900$  & 500 & Seimei (KOOLS)  \\
2019-09-25 & 58751.7 &  44.0  & $4000$--$8900$  & 500 & Seimei (KOOLS)  \\
2019-09-25 & 58751.7 &  44.0  & $4500$--$9200$  & 400 & Kanata (HOWPol) \\
2019-10-11 & 58767.9 &  60.2  & $3500$--$9000$  & 300 & HCT (HFOSC)  \\
2019-10-31 & 58787.8 &  80.1  & $3500$--$9000$  & 300 & HCT (HFOSC)  \\
2019-11-06 & 58793.6 &  85.9  & $3500$--$9000$  & 300 & HCT  (HFOSC) \\
2019-12-08 & 58825.5 & 117.8  & $3700$--$6000$  & 650 & Subaru (FOCAS) \\
2019-12-08 & 58825.5 & 117.8  & $5800$--$10000$ & 650 & Subaru (FOCAS) \\
      \hline
    \end{tabular}}\label{tb:spec_log}
    \begin{tabnote}
    \footnotemark[$\|$] Relative to the epoch of $B$-band maximum (MJD 58707.69).  \\ 
\end{tabnote}
\end{table}

\section{Results}\label{sec:results}
\subsection{Light Curves}\label{sec:lc}
\begin{figure}
 \begin{center}
  \includegraphics[width=12cm]{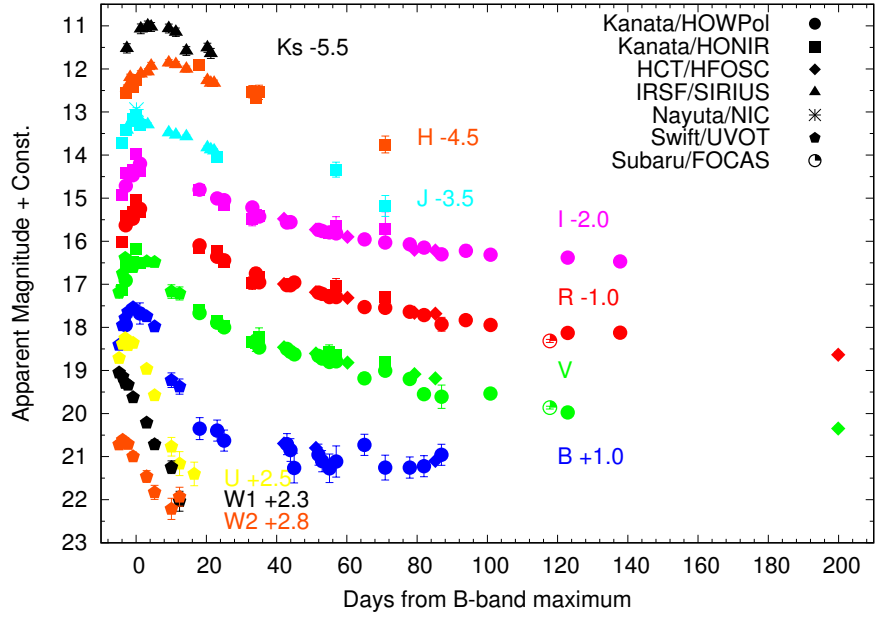} 
 \end{center}
\caption{Multi-band light curves of SN 2019muj.
The different symbols denote data that were obtained using 
different instruments (see the figure legends).
The light curve of each band is shifted vertically as 
indicated in the figure.
We adopt MJD 58707.69 $\pm$ 0.07 as 0 days.}
\label{fig:lc}
\end{figure}

\begin{figure}
\begin{center}
\includegraphics[width=12cm]{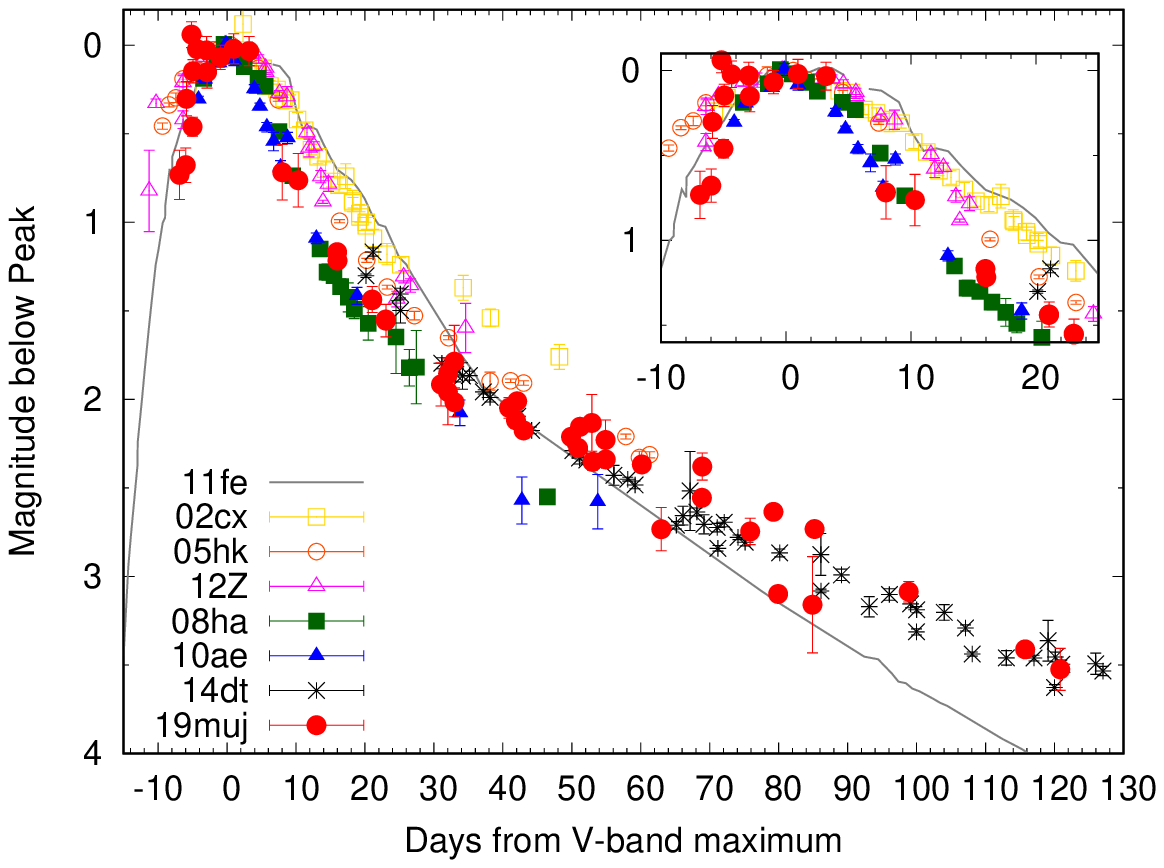}
\end{center}
\caption{$V$-band light curve of SN 2019muj. For comparison, 
we plot those of SNe 2002cx (\citealt{Li03}), 2005hk (\citealt{Sahu08}), 
2008ha (\citealt{Foley09}), 2010ae (\citealt{Stritzinger14}), 
2011fe(\citealt{Zhang16}), 2012Z (\citealt{Yamanaka15}), 
and 2014dt (\citealt{Kawabata18}).
}
\label{fig:cmp_v}
\end{figure}

Figure \ref{fig:lc} shows the multi-band light curves of SN 
2019muj, from the rising phase through the tail phase.
We compare the $V$-band light curves of SN 2019muj 
and other SNe Iax in Figure \ref{fig:cmp_v}.
We estimate the epoch of the $B$-band maximum as MJD 58707.69 
$\pm$ 0.07 (2019 Aug. 12.7) by performing a polynomial fitting to 
the data points around the maximum light.
In this paper, we refer the $B$-band maximum date as 0 days.

We derive the maximum magnitudes in the $B$ and $V$ bands as 16.623 
$\pm$ 0.025 mag and 16.380 $\pm$ 0.018 mag, respectively.
The distance modulus for VV525 is taken as $\mu = $ 32.46 $\pm$ 0.23 mag, 
which is the mean value of the results from different methods, as 
summarized in the NASA/IPAC Extragalactic Database 
(NED)\footnote{http://ned.ipac.caltech.edu/}.
The extinction through the Milky Way is estimated as $E(B-V) = 0.023$ mag 
(\citealt{Schlafly11}) for which we adopt $R_{V} = 3.1$.
We assume negligible extinction within the host galaxy, based on the 
absence of the Na D lines and the color evolution as compared with those 
of other SNe Iax.
The $B$-band absolute peak magnitude of SN 2019muj is then 
$-16.323 \pm 0.23$ mag. 
SN 2019muj is fainter by $\sim$ 1.5 mag than SNe 2002cx 
($-17.55 \pm 0.34$ mag; \citealt{Li03}) and 2005hk ($-18.02 \pm 0.32$ 
mag; \citealt{Sahu08}).
On the other hand, SN 2019muj is brighter than SNe Iax 2008ha 
($-13.74 \pm 0.15$ mag; \citealt{Foley09}, $-13.79 \pm 0.14$ 
mag; \citealt{Stritzinger14}) and 2010ae ($-13.44 \sim -15.47$ mag; 
\citealt{Stritzinger14}). 

We derive the decline rate of SN 2019muj in the $B$ and $V$ bands as 
$\Delta$m$_{15}$($B$) =  2.16 $\pm$ 0.21 mag and  
$\Delta$m$_{15}$($V$) =  1.18 $\pm$ 0.04 mag, respectively.
The $B$-band decline rate is similar to SNe 2008ha (2.17 $\pm$ 0.02 
mag; \citealt{Foley09}, 2.03 $\pm$ 0.20 mag; \citealt{Stritzinger14}) 
and 2010ae (2.43 $\pm$ 0.11; \citealt{Stritzinger14}).
This value is larger than those of the brighter SNe Iax 2002cx 
(1.29 $\sim$ 0.11 mag; \citealt{Li03}) and 2005hk (1.56 $\pm$ 0.09 mag; 
\citealt{Sahu08}).

\begin{figure}
\begin{center}
\includegraphics[width=12cm]{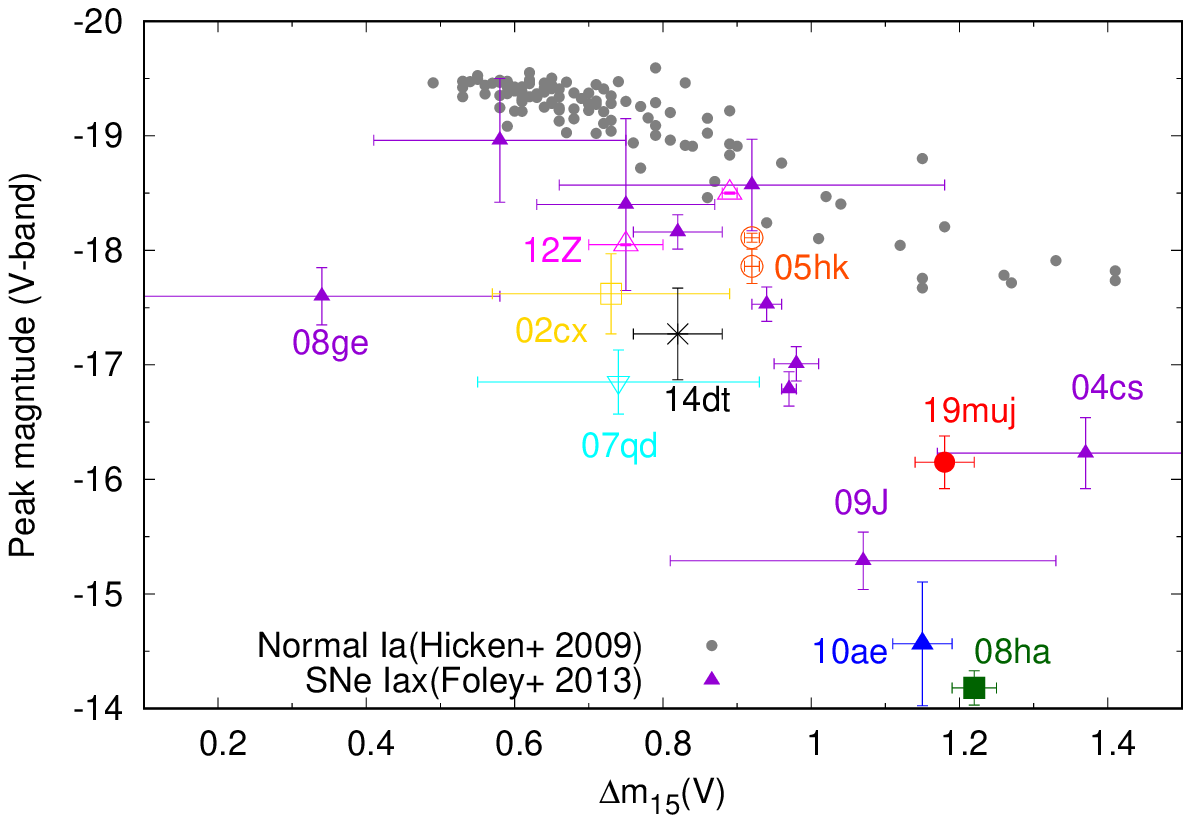}
\end{center}
\caption{Relationship of the peak absolute magnitude in the $V$ band 
with the decline rate $\Delta m_{15}$($V$).
The gray points are the data of normal SNe Ia (\citealt{Hicken09}), and 
the other points are for SNe Iax (\citealt{Li03}, \citealt{Phillips07}, 
\citealt{Sahu08}, \citealt{Foley09}, \citealt{McClelland10}, 
\citealt{Foley13}, \citealt{Stritzinger14}, \citealt{Stritzinger15}, 
\citealt{Yamanaka15}, \citealt{Kawabata18}).
For some SNe Iax, the SN IDs are given by labels.
}
\label{fig:decline}
\end{figure}

Figure \ref{fig:decline} shows the relationship between the peak 
absolute magnitude in the $V$ band and the decline rate 
$\Delta m_{15}$($V$).
The decline rate of SN 2019muj is similar to those of subluminous 
SNe Iax 2008ha and 2010ae.
The peak magnitude is however between the subluminous SNe Iax and 
bright SNe Iax. 
SN 2019muj thus belongs to the intermediate subclass between the bright 
and subluminous SNe Iax, similar to SNe 2004es and 2009J (Figure 16 of 
\citealt{Foley13}).
\citet{Barna21} also pointed out that SN 2019muj is 
located at the luminosity gap for SNe Iax.
For the intermediate SNe Iax, multi-band data covering the rising phase 
are still limited.
We discuss the light curve in the rising phase in \S\ref{sec:earlyLC}.

After $\sim$30 days, the evolution of the light curve becomes slow. 
The decline rates in $V$, $R$, and $I$-band light curves between 100 -- 
200 days are calculated 0.007 $\pm$ 0.002, 0.005 $\pm$ 0.003, 
and 0.004 $\pm$ 0.001 mag, respectively.
We compare the light curves of SN 2019muj with those of SN 2014dt 
after 60 days.
Their decline rates exhibit impressive similarity.
In \S 4, we focus on the late-phase light curves and provide 
analytical inspection.
SN 2019muj is the second case which clearly shows a slowly evolving 
light curve, being consistent with the full trapping of the 
$\gamma$-ray energy from $^{56}$Co decay (see also 
\S\ref{sec:mechanism2}).

\subsection{Spectral Evolution}\label{sec:specev}

\begin{figure}
\begin{center}
\includegraphics[width=18cm]{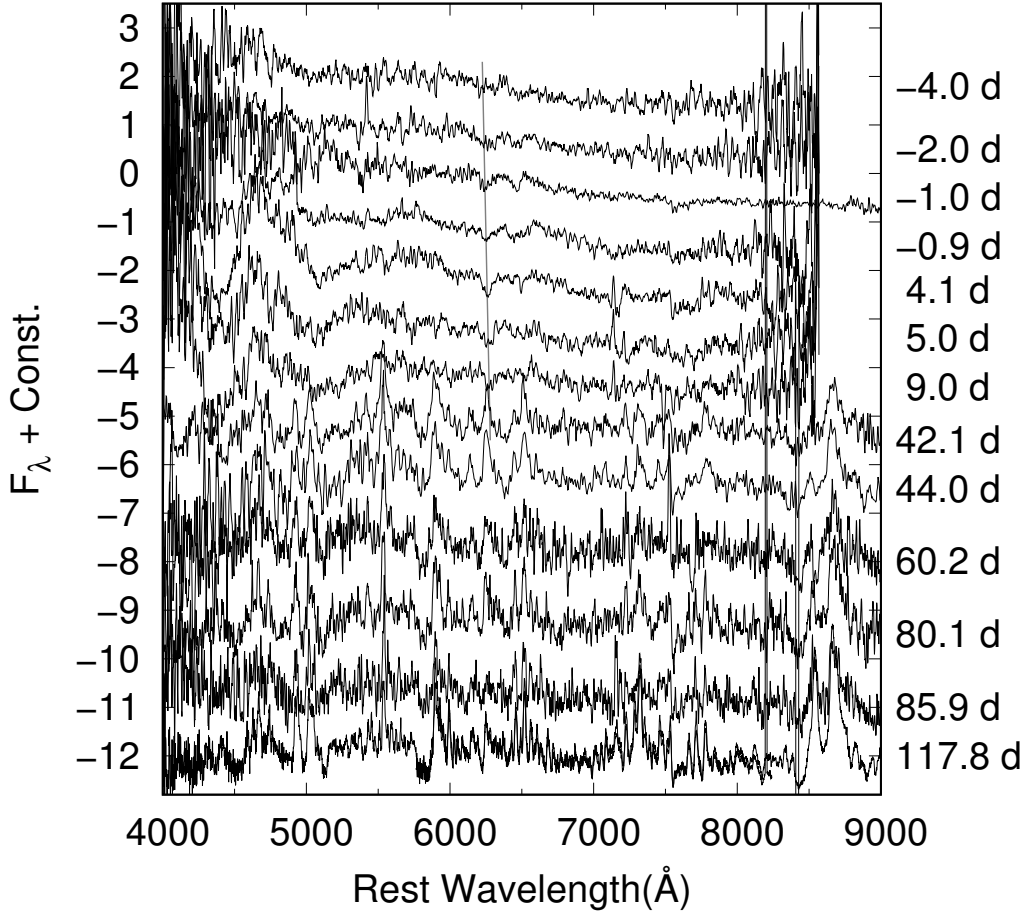}
\end{center}
\caption{Spectra evolution of SN 2019muj. The epoch of each 
spectrum is indicated on the right outside the panel.
The gray line connects the position of the Si {\sc ii} $\lambda$ 6355 line 
as a function of time.
}
\label{fig:spec}
\end{figure}

\begin{figure}
\begin{center}
\includegraphics[width=18cm]{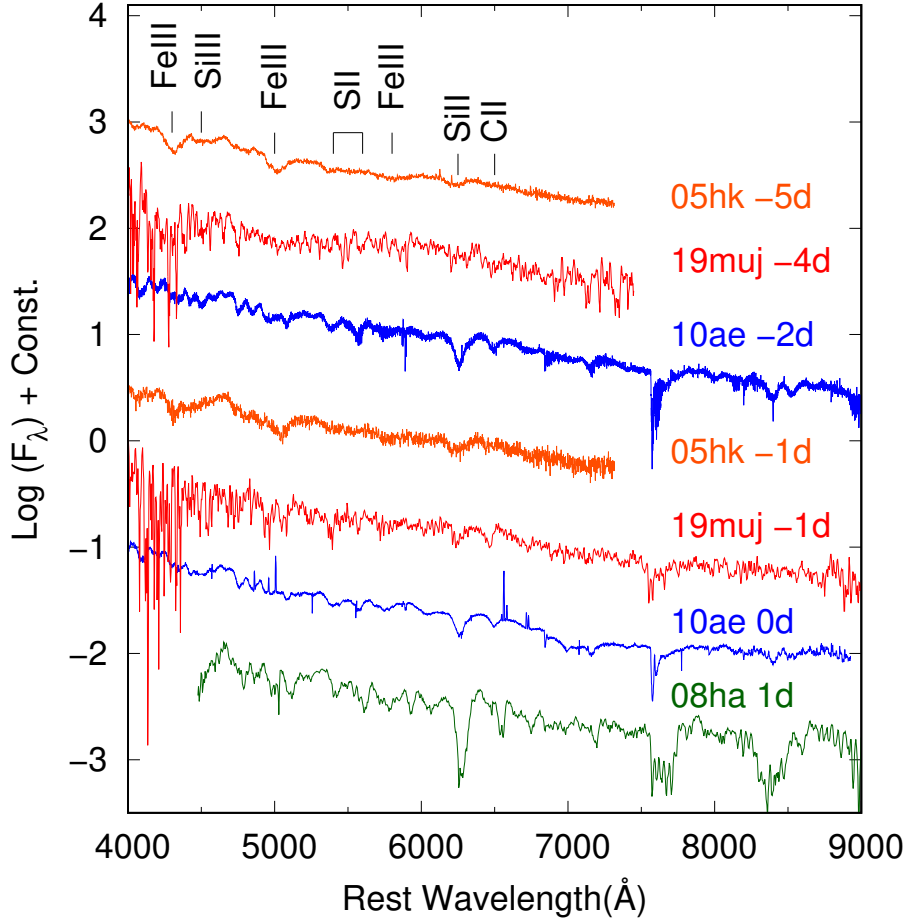}
\end{center}
\caption{Spectral comparison of SN 2019muj with other SNe Iax; 
2005hk (\citealt{Blondin12}), 2008ha (\citealt{Valenti09}), 
2010ae (\citealt{Stritzinger14}).
The phase of these spectra is from $\sim -5$ days to around the 
maximum light.
}
\label{fig:spec2}
\end{figure}

\begin{figure}
\begin{center}
\includegraphics[width=18cm]{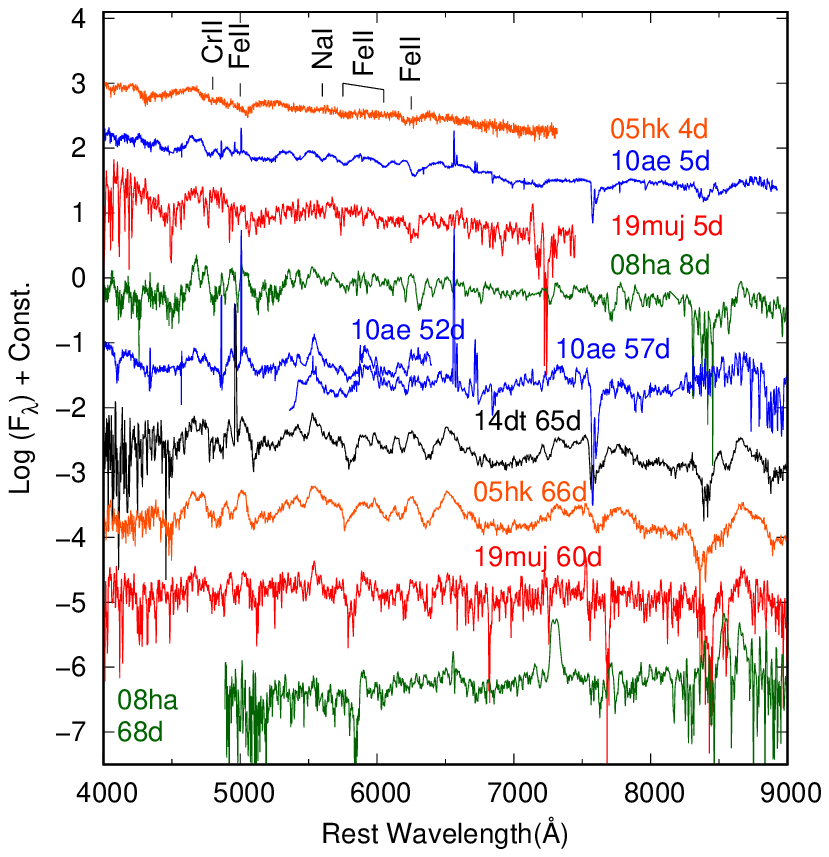}
\end{center}
\caption{Spectral comparison of SN 2019muj with other SNe Iax; 
2005hk (\citealt{Phillips07}; \citealt{Blondin12}), 2008ha 
(\citealt{Valenti09}; \citealt{Foley09}), 2010ae (\citealt{Stritzinger14}).
The phase of these spectra is from $\sim$ 5 days to $\sim$ 60 days.
}
\label{fig:spec3}
\end{figure}

\begin{figure}
\begin{center}
\includegraphics[width=18cm]{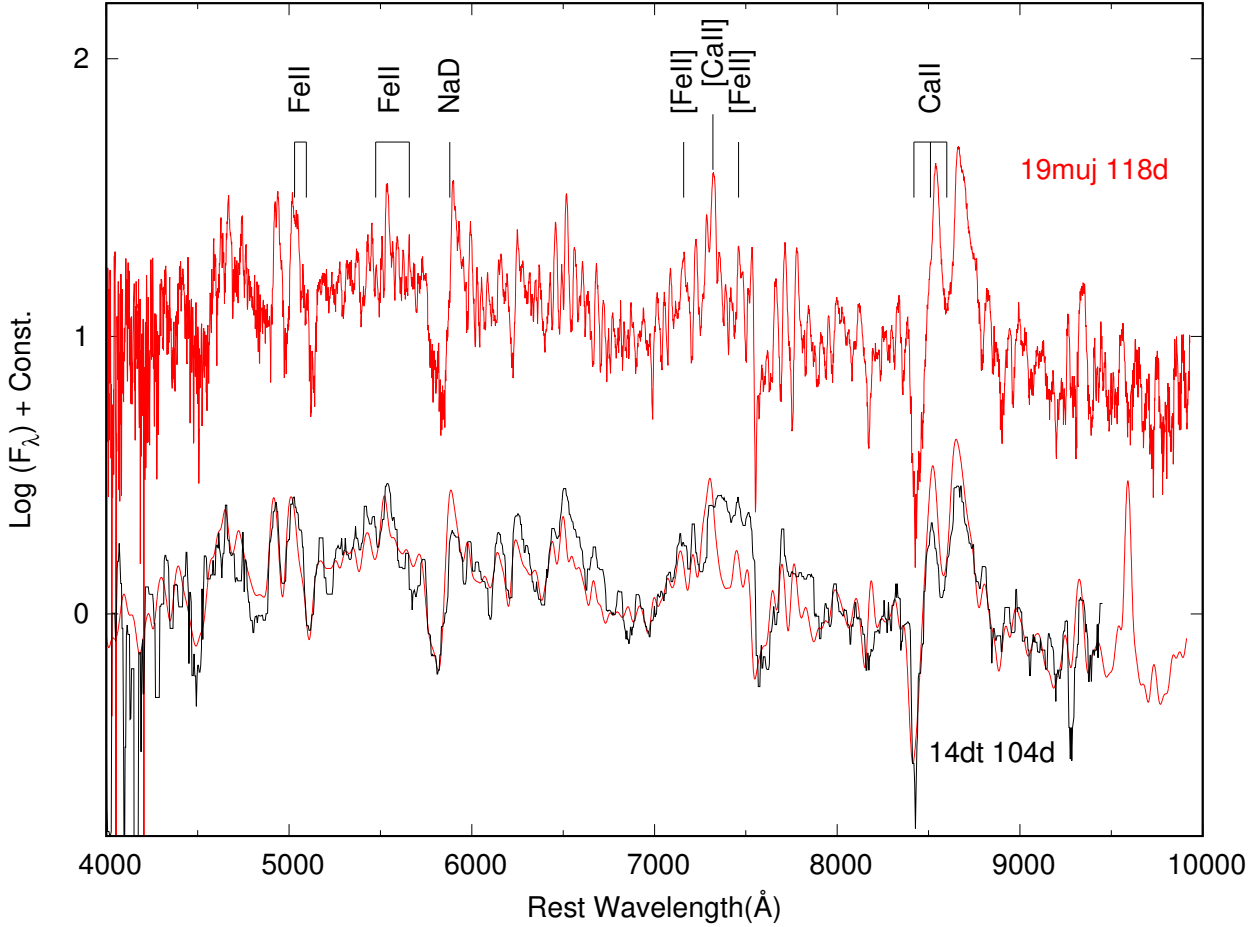}
\end{center}
\caption{Late-phase spectra of SN 2019muj (red lines) and 
SN 2014dt (black lines; \citealt{Kawabata18}). 
We overplot the Gaussian smoothed and blue-shifted spectrum of SN 2019muj on the spectrum of SN 2014dt.
}
\label{fig:spec4}
\end{figure}

We show the optical spectra of SN 2019muj from $-4.0$ days 
through $117.8$ days in Figures \ref{fig:spec} -- \ref{fig:spec4}.
In the spectra before the $B$-band maximum light, SN 2019muj shows 
a blue continuum with the narrow absorption lines of Si {\sc ii}, 
S {\sc ii}, Fe {\sc iii} and C {\sc ii} (Figures \ref{fig:spec} 
and \ref{fig:spec2}).
Although some of the spectra of SN 2019muj are somewhat noisy, 
it is clearly seen that the C {\sc ii} absorption line is as 
strong as the Si {\sc ii} $\lambda$ 6355 line during the early 
phases.
SN 2019muj is similar to SNe 2008ha and 2010ae in many spectral 
features (except for Si {\sc ii}, Ca {\sc ii} IR triplet) including 
narrow absorption lines.

After the $B$-band maximum light, SN 2019muj shows the absorption 
lines of Na {\sc i} D, Fe {\sc ii}, Fe {\sc iii}, Co {\sc ii}, and 
the Ca {\sc ii} IR triplet (Figures \ref{fig:spec} and \ref{fig:spec3}).
These absorption lines of SN 2019muj are narrow, similarly to those 
seen in the pre-maximum spectra.
SN 2008ha and SN 2010ae have similar spectra around maximum 
light (see \citealt{Stritzinger14}). 

The late phase spectrum ($\sim$ 120 days) of SN 2019muj is plotted 
in Figure \ref{fig:spec4}.
SN 2019muj shows the narrow permitted Fe {\sc ii} lines and
several forbidden lines associated with Fe, Co, and Ca {\sc ii}.
The line identification is based on \citet{Jha06} and
\citet{Sahu08}.
In this phase, there are few comparative samples of SNe Iax.
Once the spectrum of SN 2019muj is smoothed using the Gaussian
function with a kernel width of 10 pixels ($\sim 500$ km s$^{-1}$), 
and then shifted to the blue by $\sim$ 900 km s$^{-1}$ 
(Figure \ref{fig:spec4}), the similarity to SN 2014dt is striking.
The spectrum of SN 2019muj in the late phase is thus characterized
by low velocities, being consistent with its similarities to
SN 2010ae in the earlier phase.

\begin{figure}
\begin{center}
\includegraphics[width=18cm]{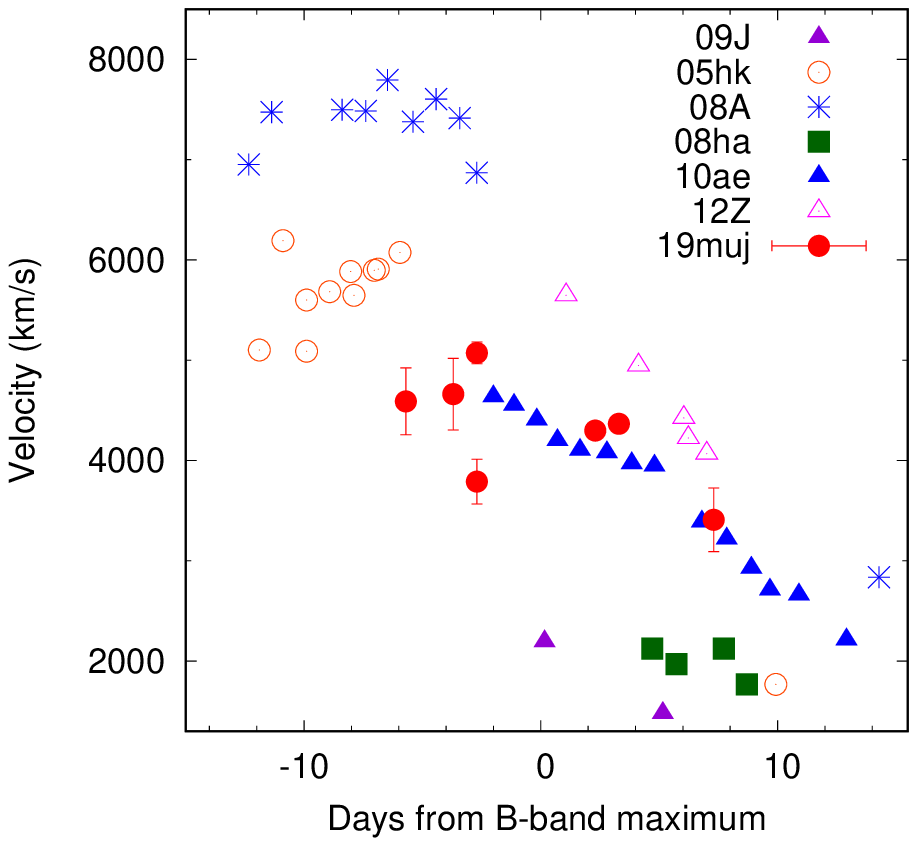}
\end{center}
\caption{Evolution of the line velocity of Si {\sc ii} $\lambda$ 6355 
in SN 2019muj. 
For comparison, we plotted those for SNe 2009J, 2005hk, 2008A, 2008ha, 
2010ae, and 2012Z.
}
\label{fig:spec5}
\end{figure}

We obtained the line velocity of Si {\sc ii} $\lambda$ 6355 by 
performing a single-Gaussian fit to the absorption line profile 
using the splot task in $IRAF$. 
We determined the uncertainty as the root mean square sum of the 
standard deviation in the fits and the wavelength resolution.
In Figure \ref{fig:spec5}, we show the velocity evolution of 
Si {\sc ii} $\lambda$ 6355.
At $-4.0$ days, the line velocity of Si {\sc ii} is 4,590 
$\pm$ 610 km s$^{-1}$.
It is slower than that of SN 2005hk ($\sim$ 6,000 km s$^{-1}$; 
\citealt{Phillips07}), while it is similar to SN 2010ae ($\sim$ 
4,500 km s$^{-1}$; \citealt{Stritzinger14}).
Around maximum light, the line velocity is 4,430 $\pm$ 750 
km s$^{-1}$, which is again similar to the velocity of 
SN 2010ae ($\sim$ 4,500 km s$^{-1}$; \citealt{Stritzinger14}). 
SN 2019muj has faster line velocities than faint SN 2008ha.

In summary, SN 2019muj shows the narrow lines and slow 
expansion velocity.
Its spectroscopic features are similar to those of subluminous 
SNe Iax such as SNe 2008ha and 2010ae.
The photometric properties of SN 2019muj are consistent with those 
of the intermediate SNe Iax. 
Note that spectra of the intermediate SNe Iax are similar to those 
of subluminous ones.

\section{Discussion}\label{sec:discussion}

SN 2019muj shows the fast decline in the early phase and the narrow
absorption lines similar to subluminous SNe Iax.
We classify it as an intermediate-luminosity SN Iax based on its 
relatively bright peak magnitude.
SN 2019muj is the second case which clearly shows a slow evolution of 
the light curve in the late phase, following bright SN Iax 2014dt
(\citealt{Kawabata18}).
Then, it is a good target for testing the weak deflagration scenario.
We also apply the analytical method to other SNe Iax, and
examine whether their light curves can be reproduced or not.

\subsection{Early Light Curves}\label{sec:earlyLC}

From the light curves covering the rising phase through the peak 
phase, we constrain the explosion date and the 
rise time in the same way as was done by \cite{Kawabata20}.
We assume the homologously expanding ``fireball model'' 
(\citealt{Arnett82}; \citealt{Riess99}; \citealt{Nugent11}) to 
estimate the explosion date. 
In this model, the luminosity/flux ($f$) increases as 
$f \propto t^{2}$, where $t$ is the time since the zero points. 
In this paper, we assume that the zero points in the time axis in 
this relation is the same for different bands (i.e., the explosion 
date), and adopt the same power-law index of 2 in all the 
$BVRI$ bands. 

\begin{figure}
\begin{center}
\includegraphics[width=12cm]{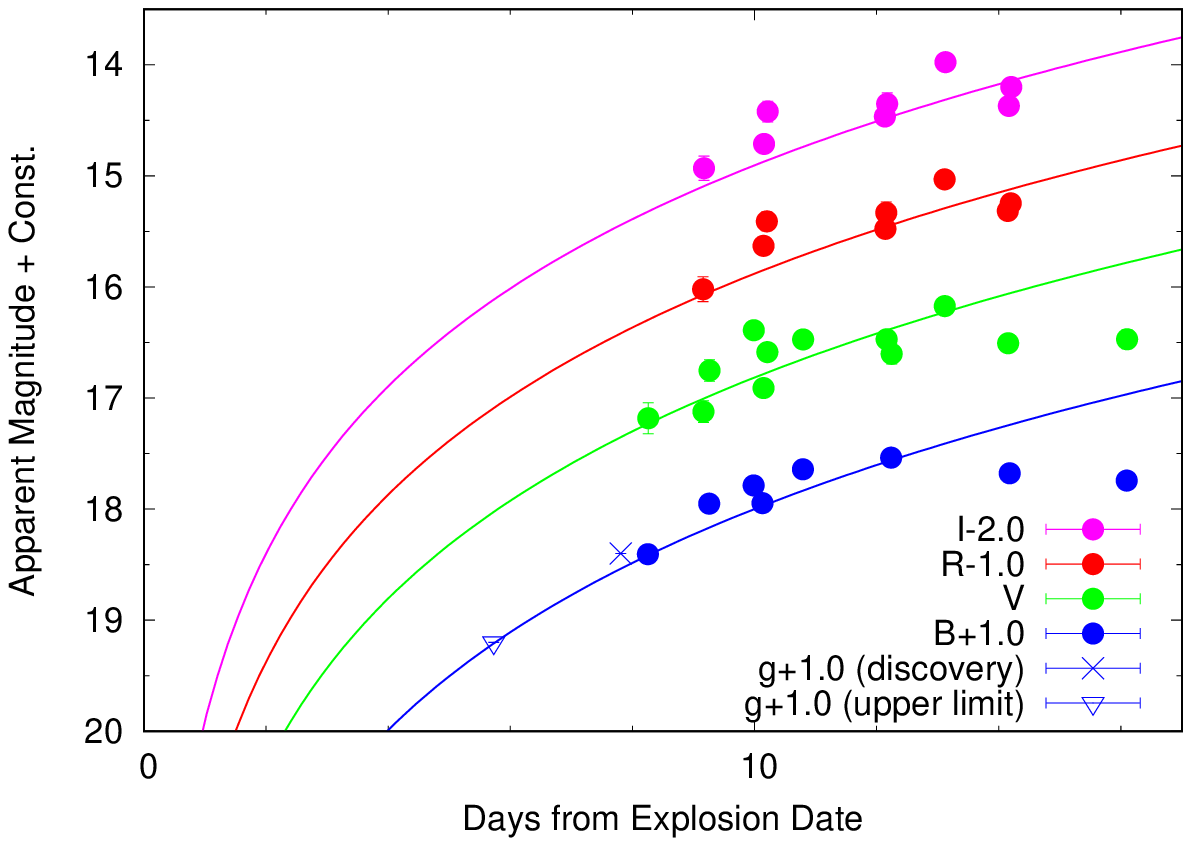}
\end{center}
\caption{Optical light curve of SN 2019muj at the early phase. 
We also plot the discovery magnitude and upper-limit magnitude 
reported by \cite{Brimacombe19}, which were obtained in the 
$g$ band. 
The explosion date is estimated to be MJD 58692.54 using the 
quadratic function (\S\ref{sec:earlyLC}).
}
\label{fig:earlyLC}
\end{figure}

In Figure \ref{fig:earlyLC}, we show the early $BVRI$-band light 
curves fitted by the fireball model only with our data. 
The rising behavior of SN 2019muj in all the bands can be 
explained by this simple fireball model. 
Through the fitting, 
we estimate the explosion date as MJD 58694.59 $\pm$ 2.1.
Note that the accuracy of the fit may depend on 
the temporal coverage of the data points used in the fit, 
and the error in the explosion date estimate above takes into 
account the change due to this effect; the  final data phase 
included in the fit is varied by a few days.
As a cross-check, we also overplotted the discovery magnitude 
and the non-detection, upper-limit magnitude.
These ASAS-SN points, as obtained in the $g$ band, were not used 
in the fit.
Given the difference of $\sim$ 0.3 mag between the $B$ and 
$g$ bands, their points are consistent with the fireball model.
No excess is found beyond the fireball prediction. 
However, it should be noted that there is not very 
much data sufficiently in the early phase to test the light curve 
behavior in the first week of the explosion.

The rise time, defined as the time interval between 
the explosion date and maximum date, may also provide some 
insights into the explosion property.
Its correlations with other observational features have been 
investigated for normal SNe Ia (e.g., \citealt{Hayden10}; 
\citealt{Ganeshalingam10}; \citealt{Jiang20}). 
This has also been investigated for SNe Iax (\citealt{Magee16}).
The rise time of SN 2019muj in the $B$ band is 13.1 $\pm$ 2.1 days.
\citet{Magee16} reported a correlation between the peak absolute 
magnitude and the decline rate either in the $R$ or $r$ band.
Because of the missing data around maximum light in the $R$ band, 
we assume that the color evolution of SN 2019muj is similar to 
that of other SNe Iax.
Then, we convert the rising time in the $B$ band to that in the  
$R$ band. 
We thereby estimate the $R$-band rise time of SN 2019muj as 16 
-- 21 days.
In Figure 5 in \cite{Magee16}, SN 2019muj is located between bright 
SNe Iax and subluminous one\textcolor{red}{s}, following the relation 
found by \cite{Magee16}.

\subsection{Explosion Parameters}\label{sec:bol}

\begin{figure}
\begin{center}
\includegraphics[width=12cm]{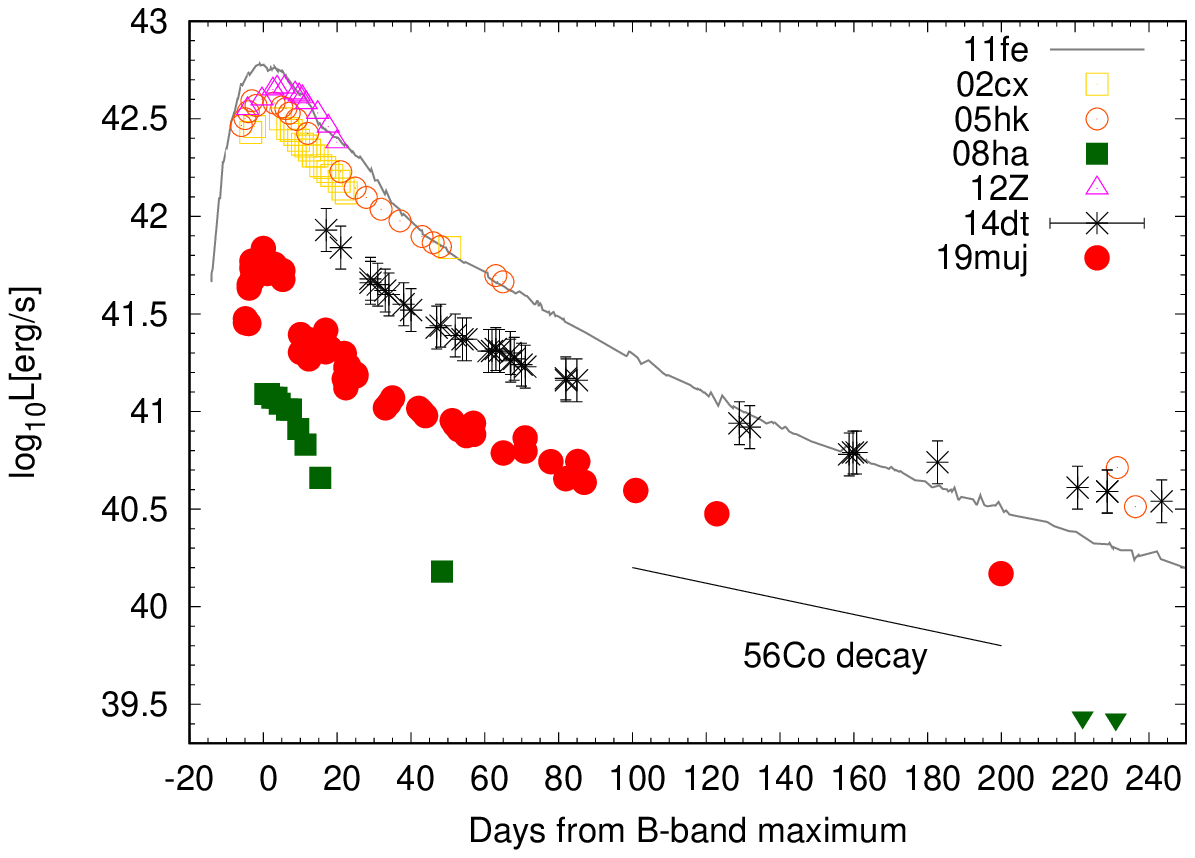}
\end{center}
\caption{Bolometric light curve of SN 2019muj. For comparison, we 
plotted the light curves of SNe 2002cx (\citealt{Li03}), 2005hk 
(\citealt{Sahu08}), 2008ha (\citealt{Foley09}; \citealt{Foley10}), 
2011fe (\citealt{Zhang16}), 2012Z (\citealt{Yamanaka15}), and 2014dt 
(\citealt{Kawabata18}).
}
\label{fig:bol}
\end{figure}

We derived the bolometric luminosity of SN 2019muj by interpolating 
the SED and integrating the fluxes within the optical bands\footnote{In 
the bolometric luminosity, the contribution from the NIR emission 
is small (SN 2012Z; \citealt{Yamanaka15}).
Then, we estimated it using only in the optical bands in this paper.}.
The derived bolometric light curve is shown in Figure \ref{fig:bol}.
SN 2019muj has intermediate luminosity among SNe Iax.
With the peak luminosity and the rising time, we estimate the 
mass of the synthesized $^{56}$Ni (\citealt{Arnett82}; 
\citealt{Stritzinger05}) as 0.01 -- 0.03 $M_{\odot}$, using the 
rise time and the peak bolometric luminosity as 13.1 $\pm$ 
2.1 days and (0.3 -- 0.8) $\times 10^{41}$ erg s$^{-1}$, respectively.
The estimated $^{56}$Ni mass is consistent with the 
value reported by \citet{Barna21}.

We then estimate $M_{\rm ej}$ and $E_{\rm k}$ of the ejecta of 
SN 2019muj by applying the scaling laws,
\begin{equation}
t_{\rm d} \propto \kappa^{1/2} \ M_{\rm ej}^{3/4} \ E_{\rm k}^{-1/4} \mbox{, and}
\label {eq:difftime}
\end{equation}
\begin{equation}
v \propto E_{\rm k}^{1/2} \ M_{\rm ej}^{-1/2} \mbox{,}
\label {eq:velocity}
\end{equation}
where $t_{\rm d}$ is the 
diffusion timescale ($\sim$ light curve width in the early phase),
$\kappa$ is the absorption coefficient for optical photons, 
and $v$ is the typical expansion velocity of the ejecta.
Here, we calibrate the relations by the properties of the 
well-studied SN Ia 2011fe (\citealt{Pereira13}) to anchor the 
normalization\footnote{The explosion parameters of SN 2011fe are 
$E_{k,11fe} = 1.2 \times 10^{51}$ erg and 
$M_{ej,11fe}$ $\sim$ 1.4 $M_{\odot}$.}.
We then obtain $E_{k,19muj}$ $\sim$ 0.02 -- 0.19 $\times$ $10^{50}$ 
erg, $M_{ej,19muj}$ $\sim$ 0.16 -- 0.95 $M_{\odot}$ and 
$\kappa_{19muj}$ / $\kappa_{11fe}$ $\sim$ 0.7 -- 4.1 (see 
\citealt{Kawabata18} for details).

\subsection{Comparison to the Weak Deflagration Model}\label{sec:mechanism2}

For SNe Iax, some explosion models have been proposed.
Among those models is the weak deflagration model leaving a bound WD 
remnant after the explosion; in this model, changing the number of 
ignition spots or the initial composition of the WD can explain the 
observational properties from bright SNe Iax to subluminous ones like 
SN 2008ha (\citealt{Fink14}; \citealt{Kromer15}).
In this section, we compare the observational properties of SN 2019muj 
to the expectations from this particular explosion model.
In the present work, we focus on the light curve 
behavior; the spectroscopic properties have been investigated by 
\citet{Barna21}, who showed that the weak/failed deflagration model can 
roughly explain the early-phase spectral evolution while the degree of 
abundance mixing found for SN 2019muj (or SNe Iax in general) is not as 
substantial as expected by the model. 

\begin{figure}
\begin{center}
\includegraphics[width=12cm]{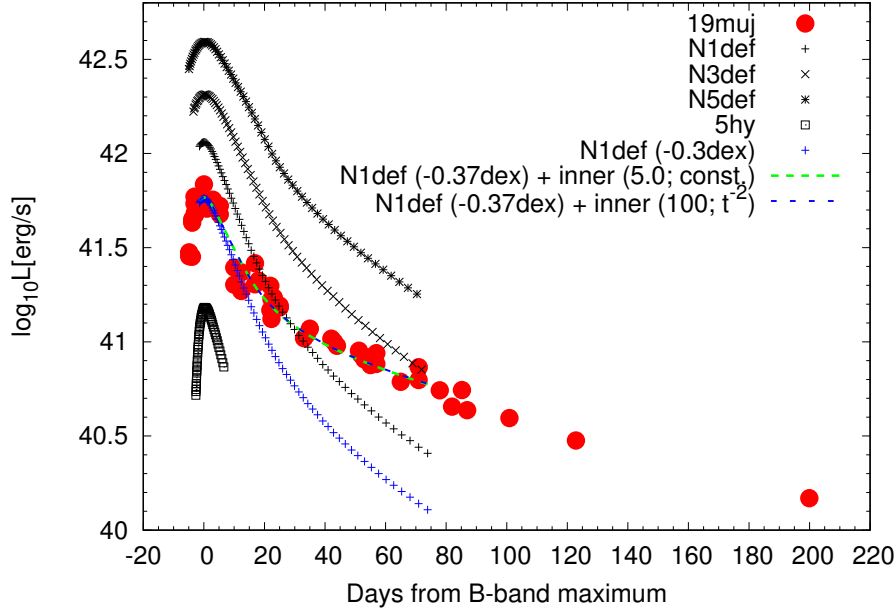}
\end{center}
\caption{Bolometric light curve of SN 2019muj. 
We compare the the weak deflagration model light curve (black points; 
\citealt{Fink14}, \citealt{Kromer15}), N5def, N3def N1def, and N5hybrid 
respectively.
The blue points are the N1def model shifted downward to fit the peak 
luminosity of SN 2019muj.
The green and blue dashed lines are the two-component model 
as the sum of the weak deflagration model and the simple radioactive-decay  
light curve model (see \S\ref{sec:mechanism2}); the evolution of the optical 
depth is taken as either $\tau =$ constant (green) or 
$\tau \propto t^{-2}$ (blue).
}
\label{fig:bol2}
\end{figure}

In the model sequence of \citet{Fink14}, the models N1def, N3def, and 
N5def cover the explosion parameters constrained for SN 2019muj 
(see \S\ref{sec:bol}).
In Figure \ref{fig:bol2}, we compare the bolometric light curve of 
SN 2019muj and the light curves calculated for these models.
The bolometric light curve of SN 2019muj is similar to that expected 
from the N1def model, while the peak luminosity of SN 2019muj is slightly 
fainter than the N1def model (see also \citealt{Barna21}).
Then, we try to match the light curve of the N1def model to that of 
SN 2019muj around the peak by shifting the model luminosity. 
This modified model can reproduce the light curve of SN 2019muj until 
$\sim$ 20 days.
However, in the tail phase, SN 2019muj shows the slow decline, unlike 
the model light curves that fade away very quickly.

To explain the light curve in the late phase, an additional energy 
source is required. 
In the model sequence by \cite{Fink14}, they indeed predict the presence 
of a bound WD remnant, which contains some $^{56}$Ni.
To test this hypothesis further, we first calculate the 
decline rate in the late phase by fitting a linear 
function to the bolometric light curve. The decline rate is found to be 
$\sim$ 0.004 dex day$^{-1}$ (0.01 mag day$^{-1}$) between 60 -- 210 days, 
which matches the expectation from nearly full-trapping of the $\gamma$-rays. 
This indicates that the additional component is powered by the decay of 
$^{56}$Co within a high-density inner component.

This additional contribution is not directly included in the model 
light curves of \citet{Fink14}. 
Indeed, the radiation process regarding this bound remnant has not been well 
understood (e.g., \citealt{Foley16}).
There are several possible scenarios in which the bound WD 
remnant would create the inner dense component: 
(1) It is possible that the component is indeed a static bound WD remnant itself. 
(2) Alternatively, the bound WD could create the inner ejecta with a slow 
expansion velocity, which may behave like the additional inner dense component 
due to its high density.
(3) Finally, there is a possibility that a high-density wind is launched 
from a bound WD remnant \citep{Foley16}.

Here, we consider a simple and phenomenological model (\citealt{Kawabata18}) 
where the power from $^{56}$Ni in the bound remnant is anyway radiated 
away through a dense environment (either in the static WD, slowly moving inner 
ejecta, or a possible wind), without much energy loss nor time delay.
As the bolometric luminosity in the late phase, we thus consider the 
simple radioactive-decay light curve model (\citealt{Maeda03}).
\begin{eqnarray}
L_{\rm bol} &=& M(^{56}\mbox{Ni})_{in} \left[ e^{(-t/8.8{\rm\ d})} \epsilon_{\gamma,\rm Ni} (1 - e^{-\tau}) \right. \nonumber \\
& & + \left. e^{(-t/111{\rm\ d})} \left\{ \epsilon_{\gamma,\rm Co} (1 - e^{-\tau}) + \epsilon_{e^{+}} \right\} \right] 
\label{eq:one-l}
\end{eqnarray}
where $\epsilon_{\gamma, \rm Ni} = 3.9 \times 10^{10}$ erg s$^{-1}$ g$^{-1}$
is the energy deposition rate by $^{56}$Ni via $\gamma$-rays,
$\epsilon_{\gamma, \rm Co} = 6.8 \times 10^{9}$ erg s$^{-1}$ g$^{-1}$ 
and $\epsilon_{e^{+}} = 2.4 \times 10^{8}$ erg s$^{-1}$ g$^{-1}$
are those by the $^{56}$Co decay via $\gamma$-rays and positron 
ejection.
The optical depth of the inner component, $\tau$, will 
evolve with time differently for the different scenarios for the inner 
component; this is expected to be constant for the bound WD, or to evolve 
as $\tau \propto t^{-2}$ for the slowly moving ejecta. 
These two possibilities will be examined below. 
In the case of the WD wind, the evolution is not trivial, but it likely 
decreases with time as the mass-loss rate goes down; it is probably a 
reasonable guess that the rate of the decrease for the wind case is between 
the two cases mentioned above.
Note that $M$($^{56}$Ni)$_{\rm in}$ and the optical depth 
here refer to those of the inner component alone without the outer component 
estimated in \S\ref{sec:bol}.

We then add this additional component to the model light curve of 
\cite{Fink14}.
In Figure \ref{fig:bol2}, we show this combined light curve model. 
The mass of $^{56}$Ni in the inner component is derived as $\sim$ 0.018 M$_{\odot}$. 
For the case of the constant optical depth, we have derived $\tau$ $\sim$ 5.0.
$\tau$ is the lower limit, and any value larger than this does not 
affect the change in the bolometric luminosity.
For the case of the slow-moving inner ejecta, the optical depth is 
characterized as follows; 
\begin{equation}
\tau \simeq 1000 \times \left[\frac{(M_{\rm ej}/M_{\odot})^{2}}{E_{51}}\right]_{in} \left(\frac{t}{\mbox{day}}\right)^{-2} \mbox{ ,}
\label{eq:tau}
\end{equation}
$M_{\rm ej}$ is the ejecta mass, and $E_{51}$ is the kinetic energy 
in $10^{51}$ erg (both only for the inner component).
We can reproduce the light curve with  
[$(M_{\rm ej}/M_{\odot})^{2}/{E_{51}}$]$_{in}$ $\sim$ 100. 
The analysis here indicates the existence of the inner dense component, 
irrespective of the specific scenarios for its origin.
The high-density component in the inner part was also seen in the bright 
SN Iax 2014dt thought similar analysis (\citealt{Kawabata18}); we thus 
find this inner component as a common property shared by bright and 
intermediate luminosity SNe Iax.

As shown above, the light curve is well reproduced irrespective of the evolution 
of the optical depth. 
In other words, from this analytical model of the bolometric light curve, it is 
difficult to distinguish the scenarios for the origin of the inner component.
This stems from the observed decay rate following the full trapping of the 
$\gamma$-rays; as long as $\tau \gg 1$, the result is not dependent on the 
time evolution of the optical depth. 
Indeed, in the model for the slowly moving ejecta, the optical depth will 
decrease below unity $\sim$ one year after the explosion. 
Therefore, further following the light curve evolution even longer than 
the observation in the present work may be useful to distinguish at least the 
slowly moving ejecta scenario. 
Also, the spectral evolution in the late phase is a key to addressing this question.

\subsection{Bound WD Remnants in all SNe Iax?}\label{sec:mechanism}

In this section, we expand the analysis in \S \ref{sec:mechanism2} on 
the light curve behavior to a sample of SNe Iax.
We estimate the bolometric luminosities of these SNe Iax using 
the same method as adopted for SN 2019muj (see Appendix for further 
details).
In Figure \ref{fig:km2_1}, \ref{fig:km2_2}, \ref{fig:km2_3} and \ref{fig:km2_4}, 
we compare the bolometric light curves of 
SNe Iax and those predicted by \citet{Fink14} for the weak deflagration 
model series. 
We can find a subset of the model series that can fit the early phase 
light curve of individual SNe Iax. 
However, the models can not replicate the late-phase slow evolution seen 
in the observed light curves, as is similar to the case for 
SN 2019muj, except for SNe 2011ay and 2019gsc.
Since their models do not consider the inner-part ejecta components, 
we further consider the light curve model with a variety of the inner-part 
ejecta properties (see \S \ref{sec:mechanism2}) to fit the bolometric 
light curves of other SNe Iax.
Using the same methodology as in \S\ref{sec:mechanism2}, 
we try to fit the combined light curve model with the two cases where $\tau$ 
is either constant (i.e., the static remnant) or evolve as $\tau \propto t^{-2}$ 
(i.e., the slowly moving inner ejecta).
Their light curves are well reproduced by this model sequence 
(Figure \ref{fig:km2_1}, \ref{fig:km2_2}, \ref{fig:km2_3}, and \ref{fig:km2_4}).
Again, similarly to the case for SN 2019muj, we unfortunately 
can not constrain the time evolution of the optical depth with the available 
data set. 
This highlights the importance of observations in even later phases 
(\S \ref{sec:mechanism2}).

Most of SNe Ia are found to have a high-density component 
in the innermost layer according to the above analyses.
Extremely long-term photometric observations are available for a few 
SNe Iax (SN 2008ha; \citealt{Foley14}, SN 2012Z; \citealt{McCully21}), which show 
slowly decaying light curves even in the extremely late phases ( $\geq$ 1400 days).
These observations are consistent with our analysis, and probably support the 
presence of the inner component.
The very slow decay even in such a late phase may disfavor the possibility of 
the homologously expanding inner ejecta as an origin of the inner-dense component. 
However, the luminosity in the very late phase may indeed decrease even more slowly 
than the radioactive decay, which might require an additional power source, e.g., 
CSM interaction \citep{McCully21}; then it wold not provide a strong constraint on 
the evolution of the optical depth of the inner component. 
Further observation and analysis covering the full evolution of SNe Iax will help 
understand their nature.

The explosion simulations by \citet{Fink14} predict the $^{56}$Ni 
masses in the inner and outer layers for a given set of input parameters 
to characterize the explosion. 
In Table \ref{tb:datasets}, we show the parameters obtained by our fit 
to the observed light curves.
The $^{56}$Ni masses required to fit the late-phase light curves are 
substantially larger than the model predictions.
Especially, SNe 2009ku and 2008ge show large discrepancies.

An additional issue is that we do not see a correlation between 
the properties of the outer ejecta and the inner component 
expected in the explosion simulations. 
Figure \ref{fig:bol3} shows the relationship between the obtained 
$^{56}$Ni mass in the outer layer and that in the inner layer.
\citet{Fink14} predicted that the model with a larger amount of 
$^{56}$Ni in the outer ejecta has less $^{56}$Ni mass in the inner 
layer.
However, our results do not follow a clear correlation.
This discrepancy indicates that there may still be missing functions 
in the explosion mechanism of SNe Iax and the origin(s) of their diverse 
properties that we have not considered in our current analysis. 

We also check if there is a relation between the 
ejecta mass in the outer layer and the optical depth in the inner component. 
In the model calculation by \citet{Fink14}, for a smaller explosion energy, 
the bound WD mass becomes larger and the outer ejecta mass becomes smaller. 
The combination of the larger WD mass and smaller explosion energy likely 
leads to a larger optical depth of the inner component along with this scenario, 
while the detail will be dependent on the specific origin for the inner component. 
We may anyway expect, at least roughly, an anti-correlation between the outer 
ejecta mass and the inner optical depth. 
We show this exercise in Figure \ref{fig:km2_5}. 
We do not see a clear relation between the two parameters. 
Similarly to the conclusion reached by the non-correlation between the $^{56}$Ni 
masses in the inner and the outer components, the finding here would further 
indicate that an important function is still missing in our analysis or in our 
understanding of the nature of SNe Iax.

\begin{figure*}
\begin{center}
 \begin{minipage}{0.44\hsize}
   \includegraphics[width=7.8cm]{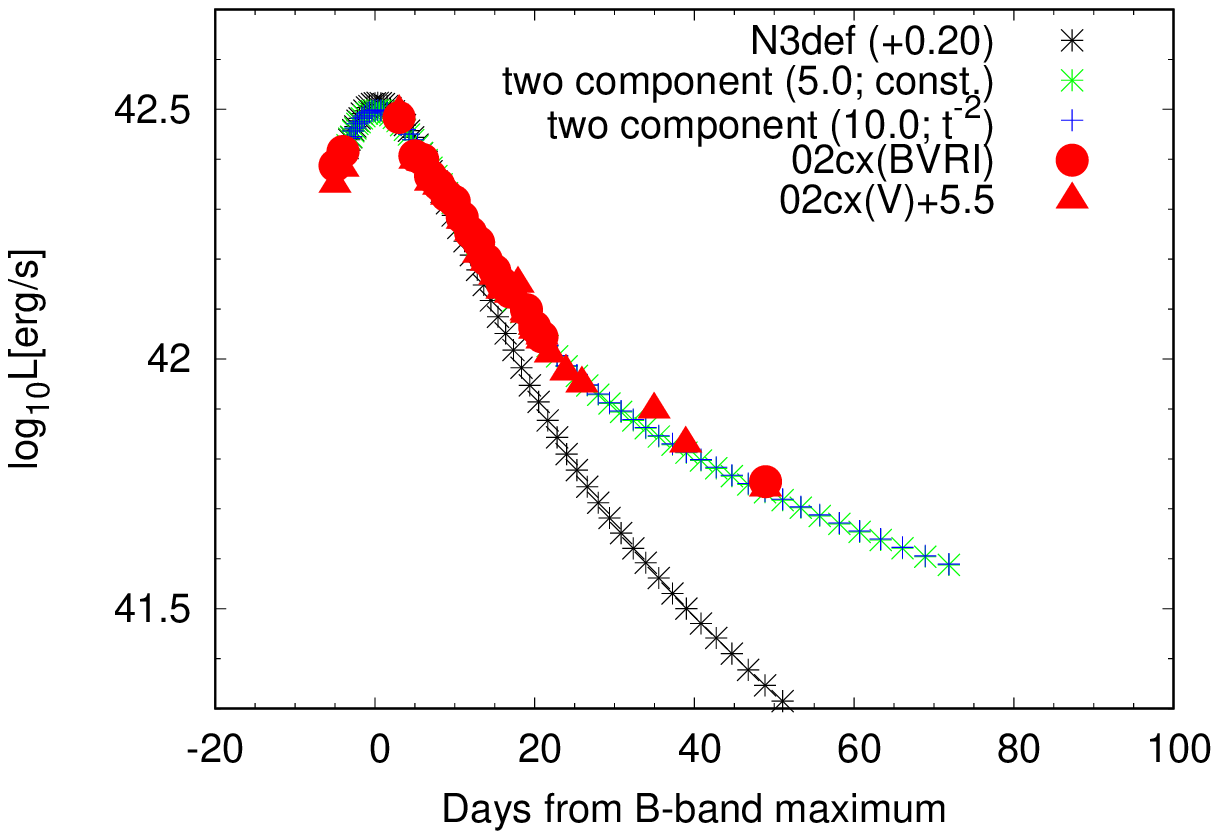}
 \end{minipage}
 \begin{minipage}{0.44\hsize}
   \includegraphics[width=7.8cm]{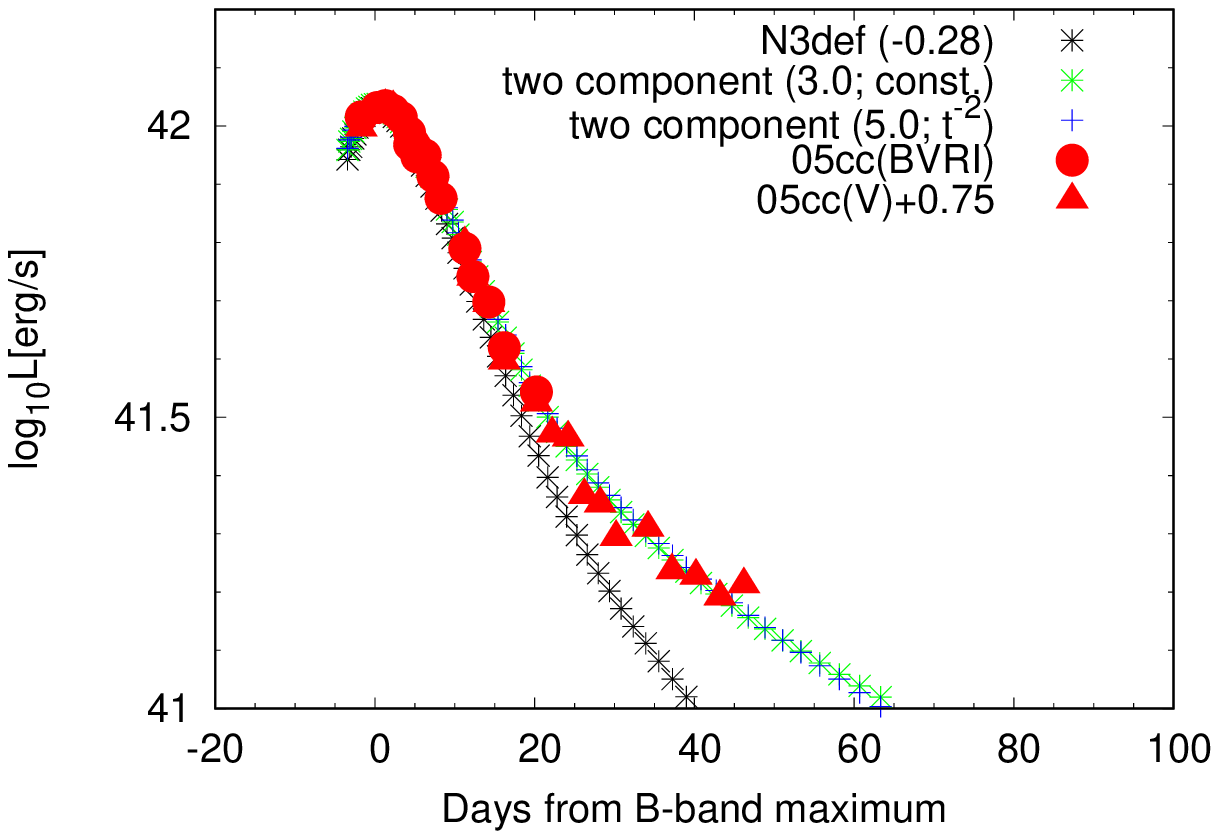}
 \end{minipage}
  \begin{minipage}{0.44\hsize}
   \includegraphics[width=7.8cm]{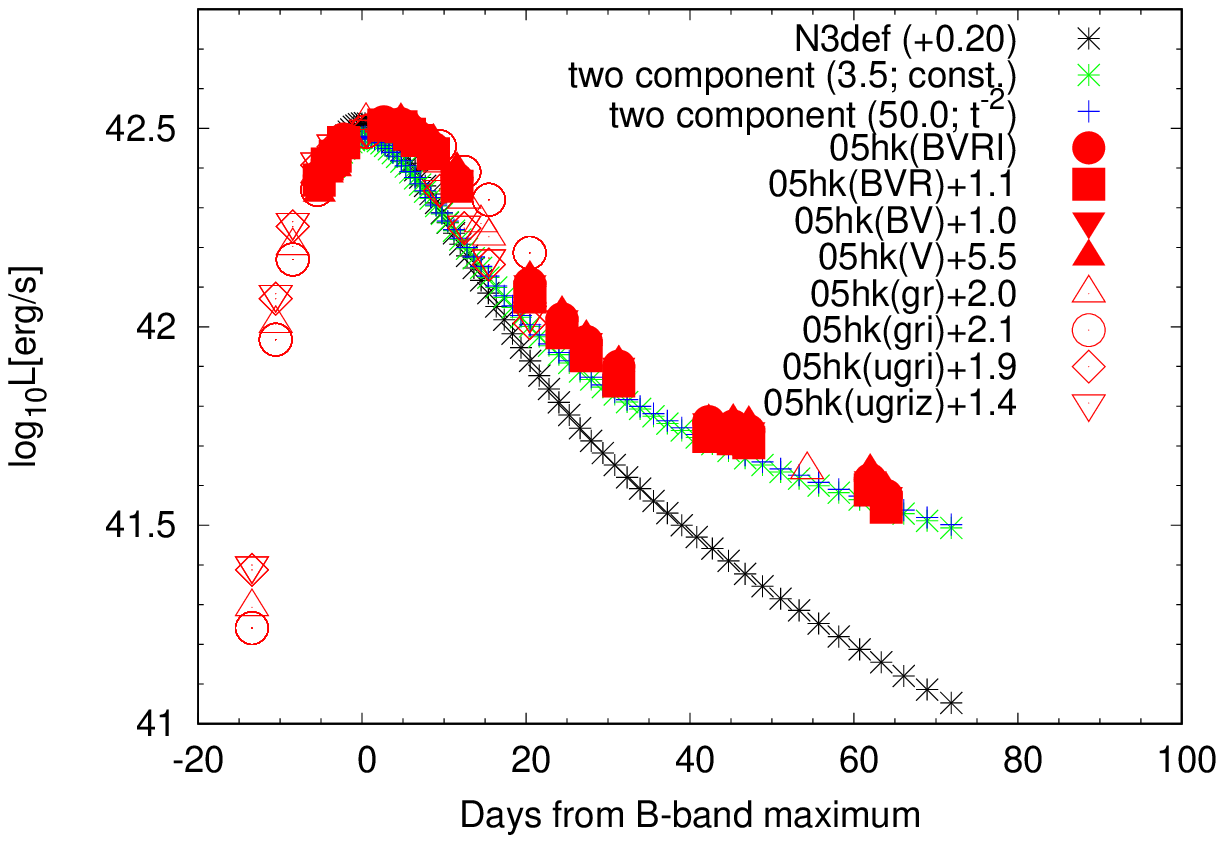}
 \end{minipage}
   \begin{minipage}{0.44\hsize}
   \includegraphics[width=7.8cm]{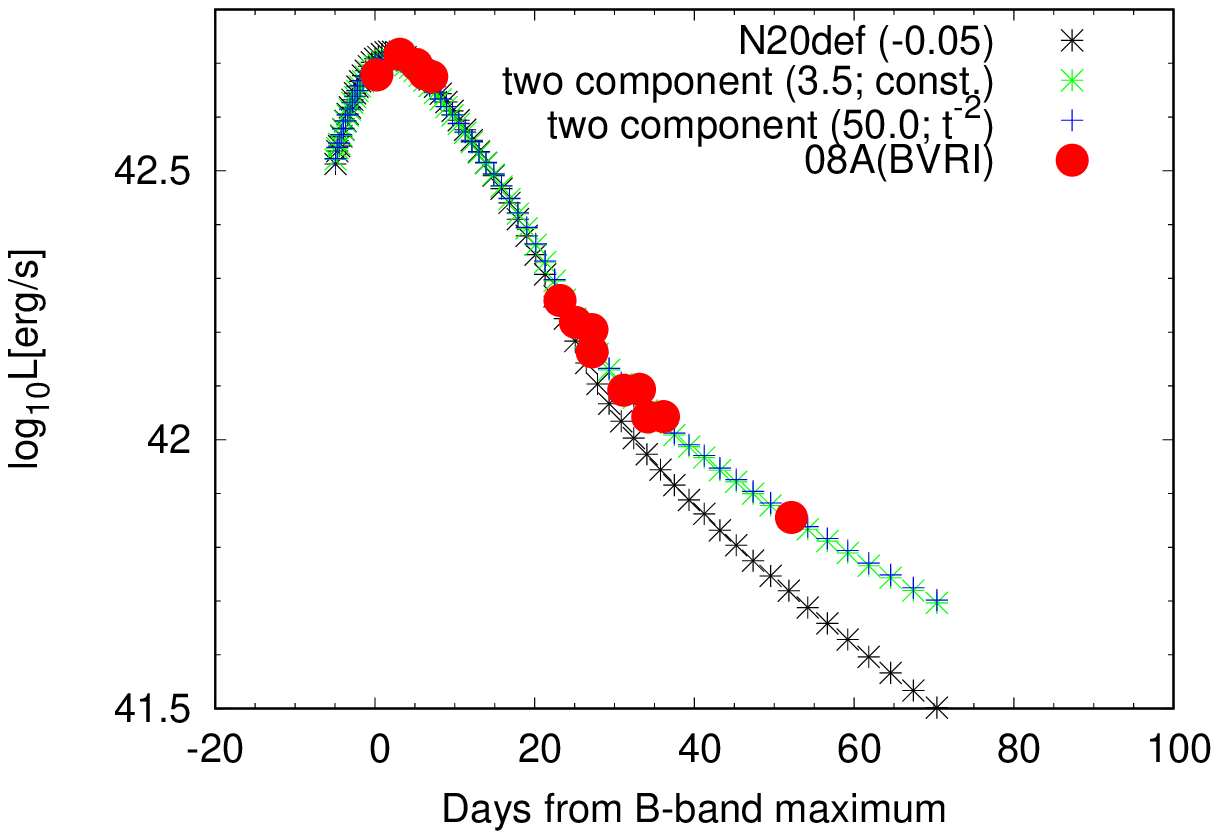}
 \end{minipage}
\end{center}
\caption{Bolometric light curves of SNe Iax (red points). 
We compare the weak deflagration model light curve (black 
points; \citealt{Fink14}).
The black points are shifted to fit the peak luminosity of each SN Iax.
The green and blue points are the two-component model 
as the sum of the weak deflagration model and the simple radioactive-decay  
light curve model (see \S\ref{sec:mechanism2}); the evolution of the 
optical depth is taken as either $\tau =$ constant (green) or 
$\tau \propto t^{-2}$ (blue).
}
\label{fig:km2_1}
\end{figure*}

\begin{figure*}
\begin{center}
 \begin{minipage}{0.44\hsize}
   \includegraphics[width=7.8cm]{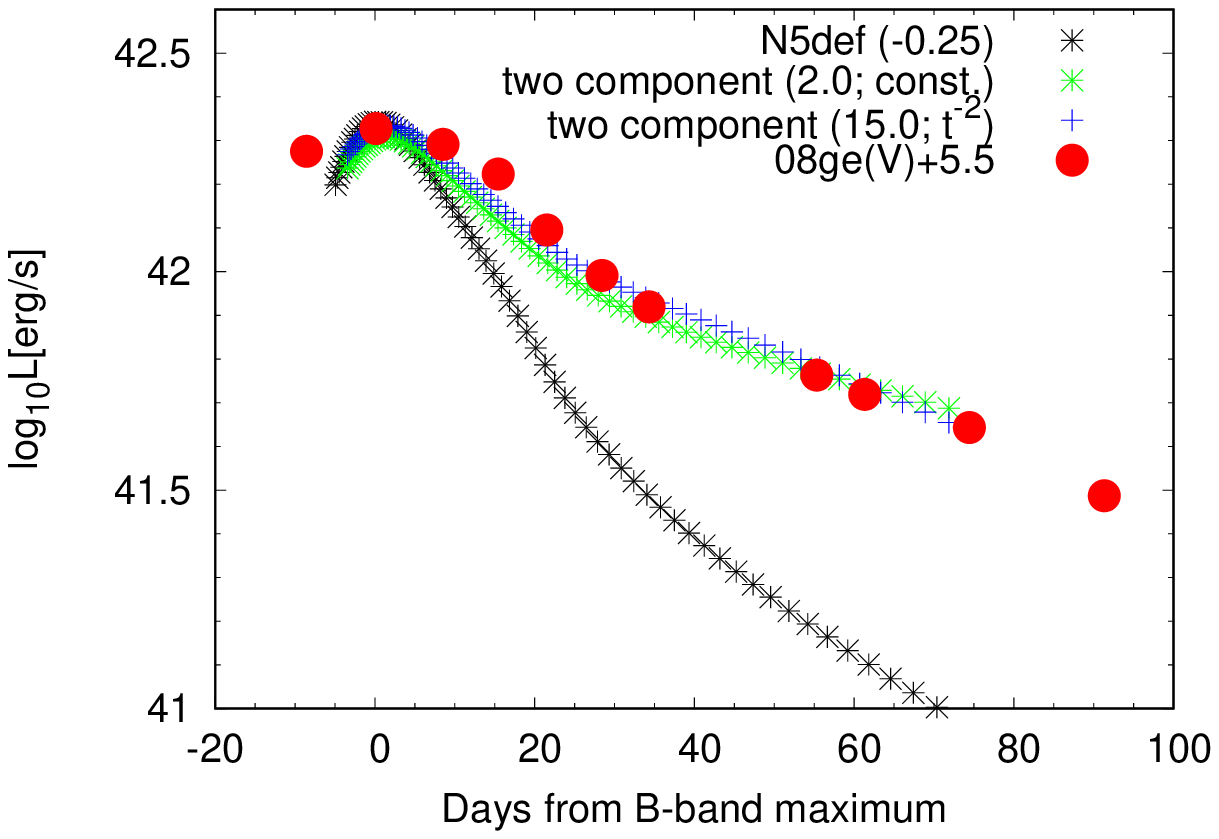}
 \end{minipage}
 \begin{minipage}{0.44\hsize}
   \includegraphics[width=7.8cm]{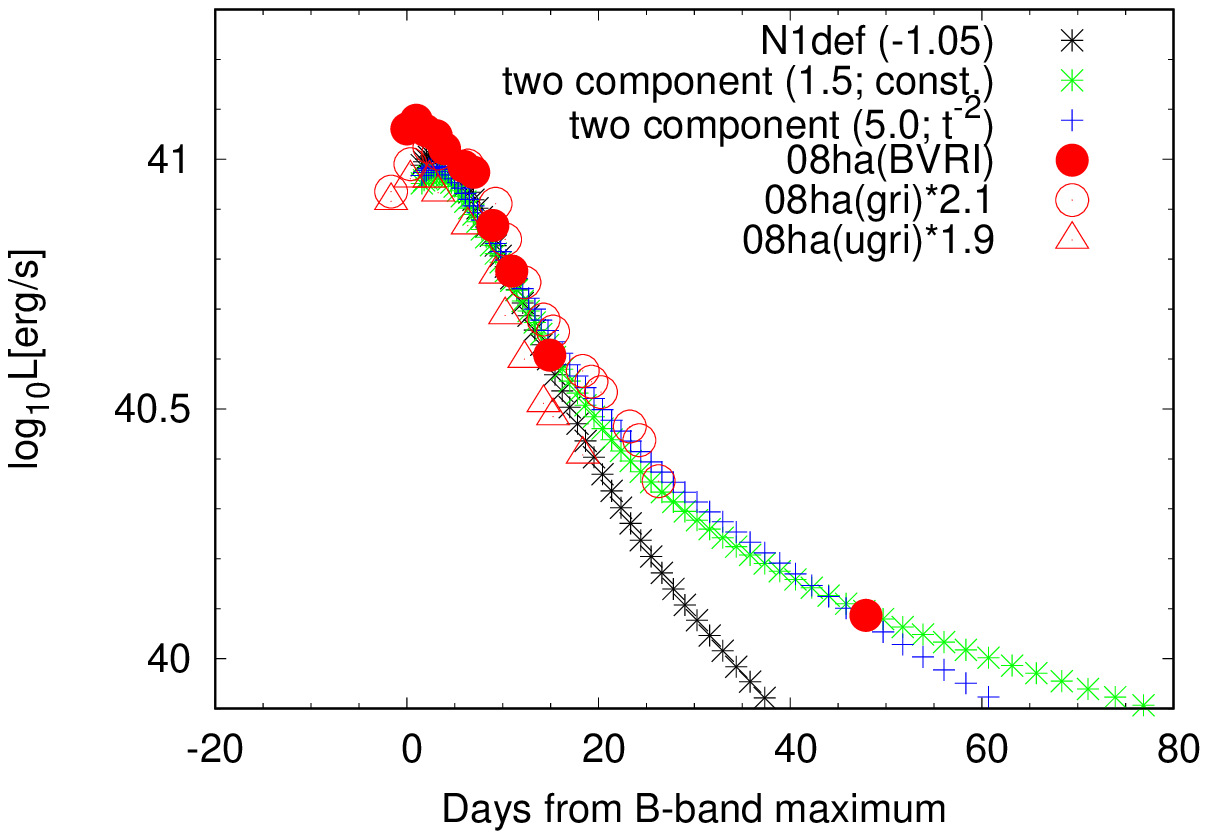}
 \end{minipage}
  \begin{minipage}{0.44\hsize}
   \includegraphics[width=7.8cm]{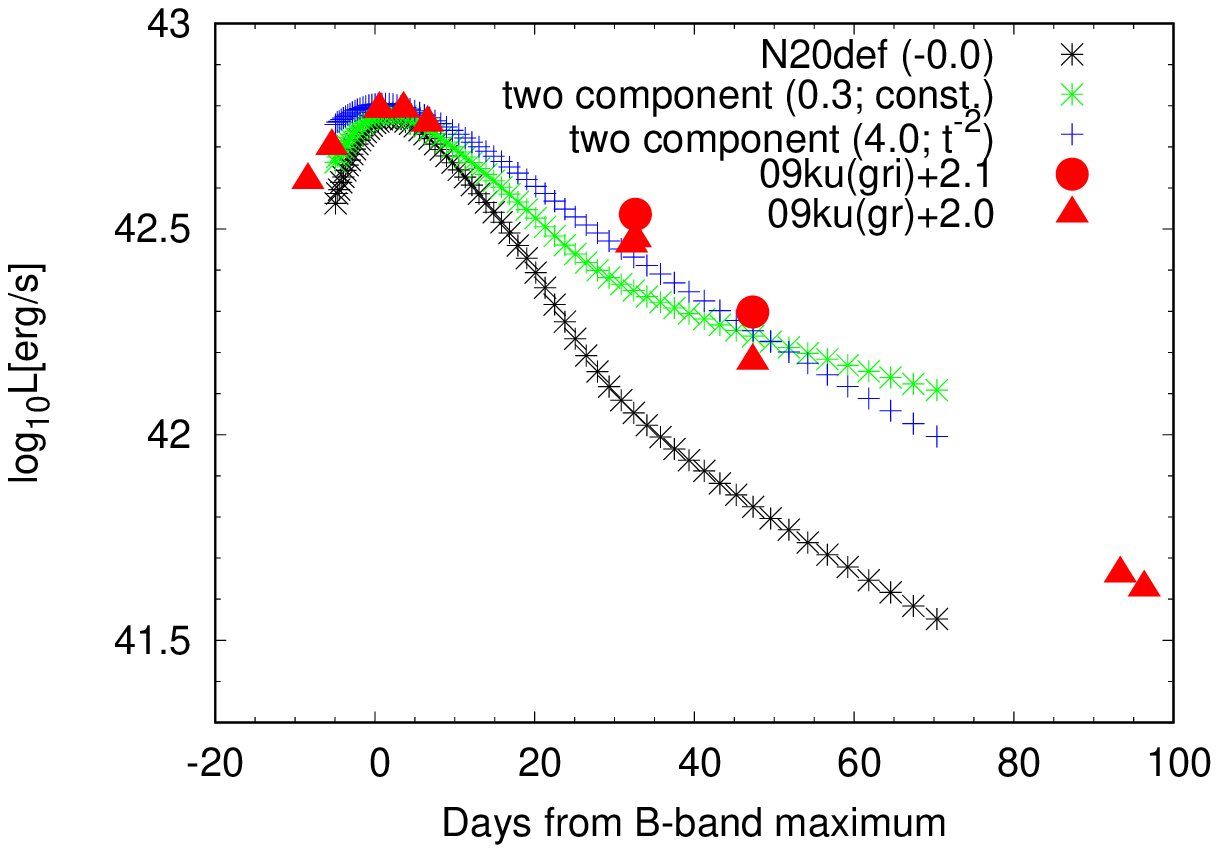}
 \end{minipage}
   \begin{minipage}{0.44\hsize}
   \includegraphics[width=7.8cm]{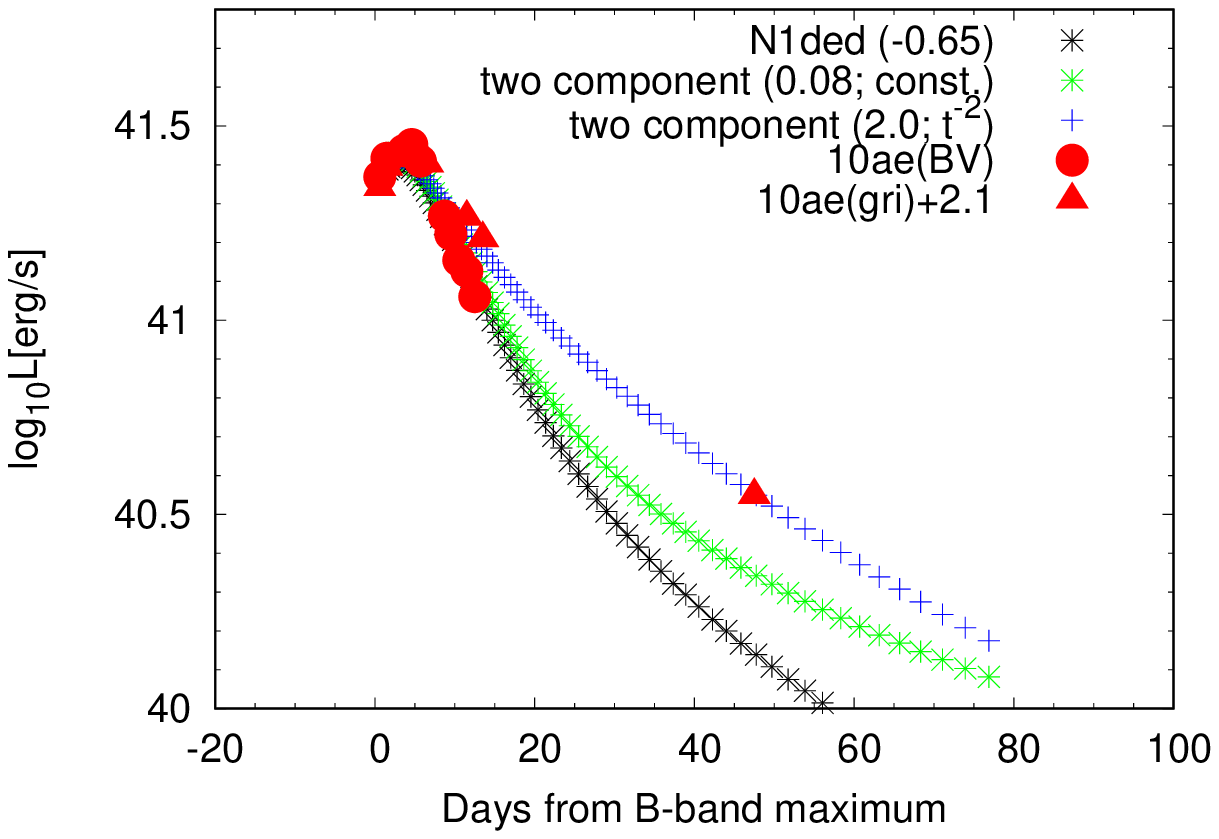}
 \end{minipage}
\end{center}
\caption{The same as Figure \ref{fig:km2_1} but for different SNe Iax.
}
\label{fig:km2_2}
\end{figure*}

\begin{figure*}
\begin{center}
 \begin{minipage}{0.44\hsize}
   \includegraphics[width=7.8cm]{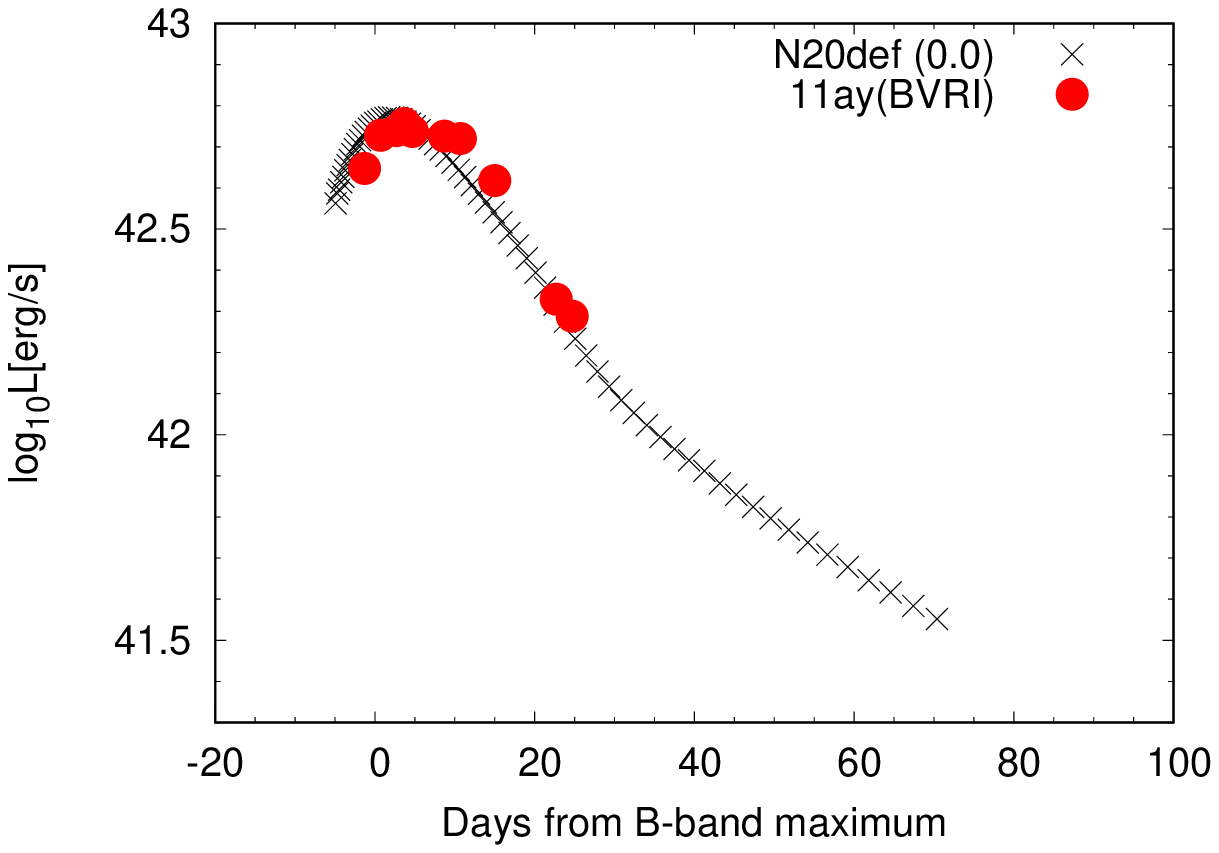}
 \end{minipage}
 \begin{minipage}{0.44\hsize}
   \includegraphics[width=7.8cm]{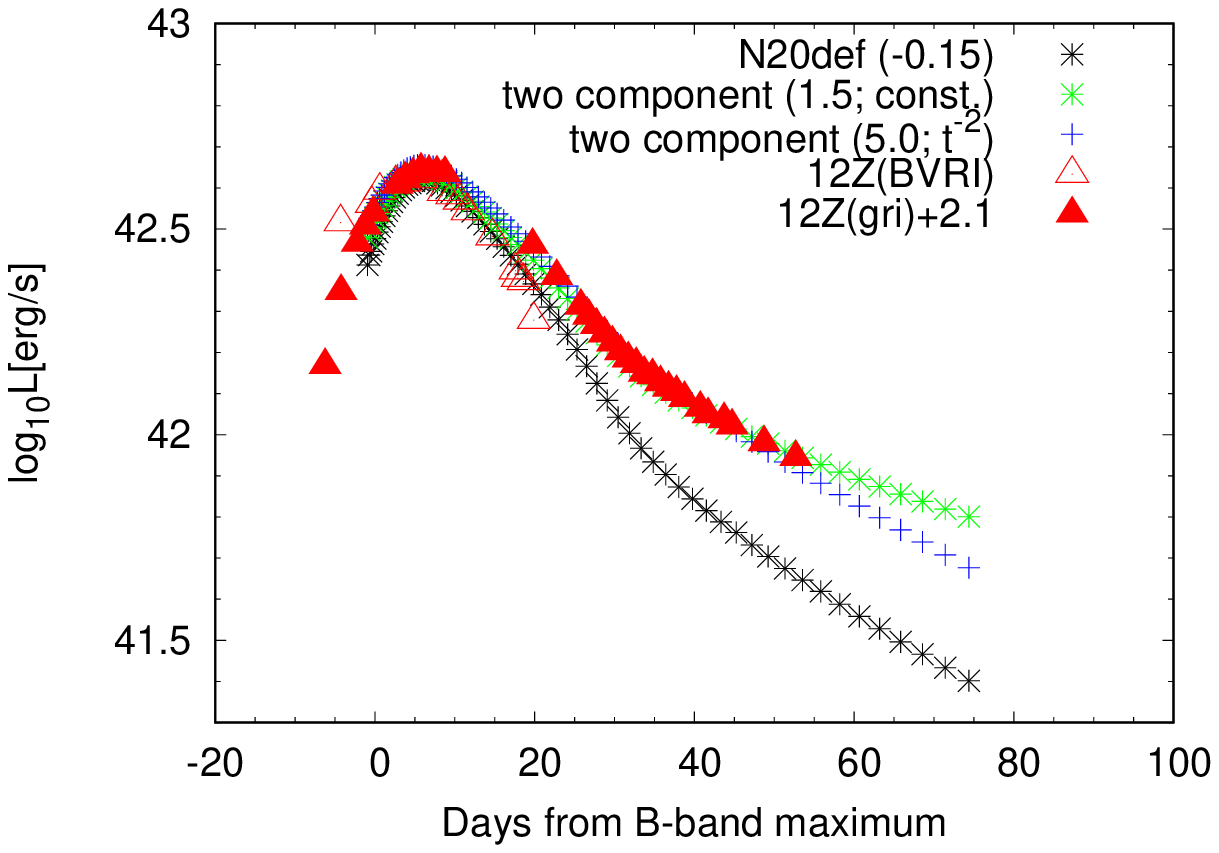}
 \end{minipage}
  \begin{minipage}{0.44\hsize}
   \includegraphics[width=7.8cm]{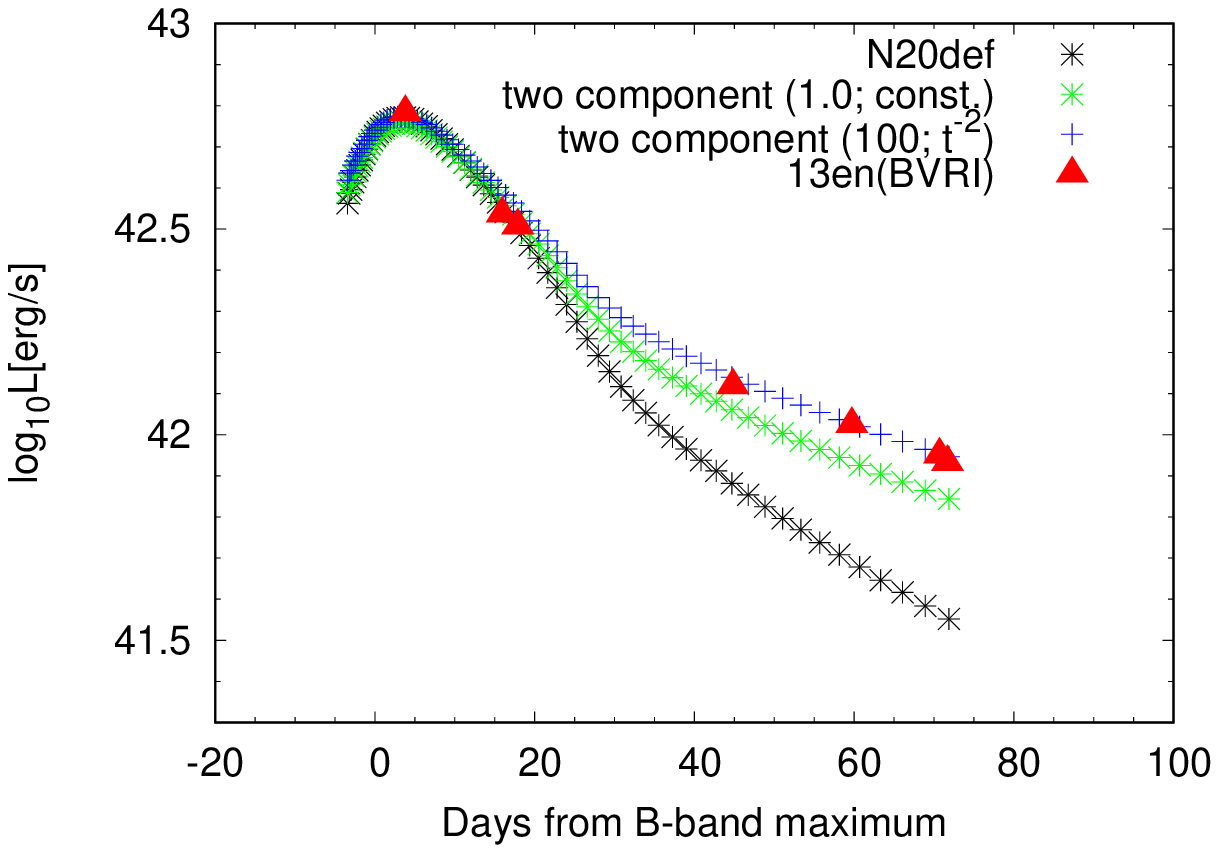}
 \end{minipage}
   \begin{minipage}{0.44\hsize}
   \includegraphics[width=7.8cm]{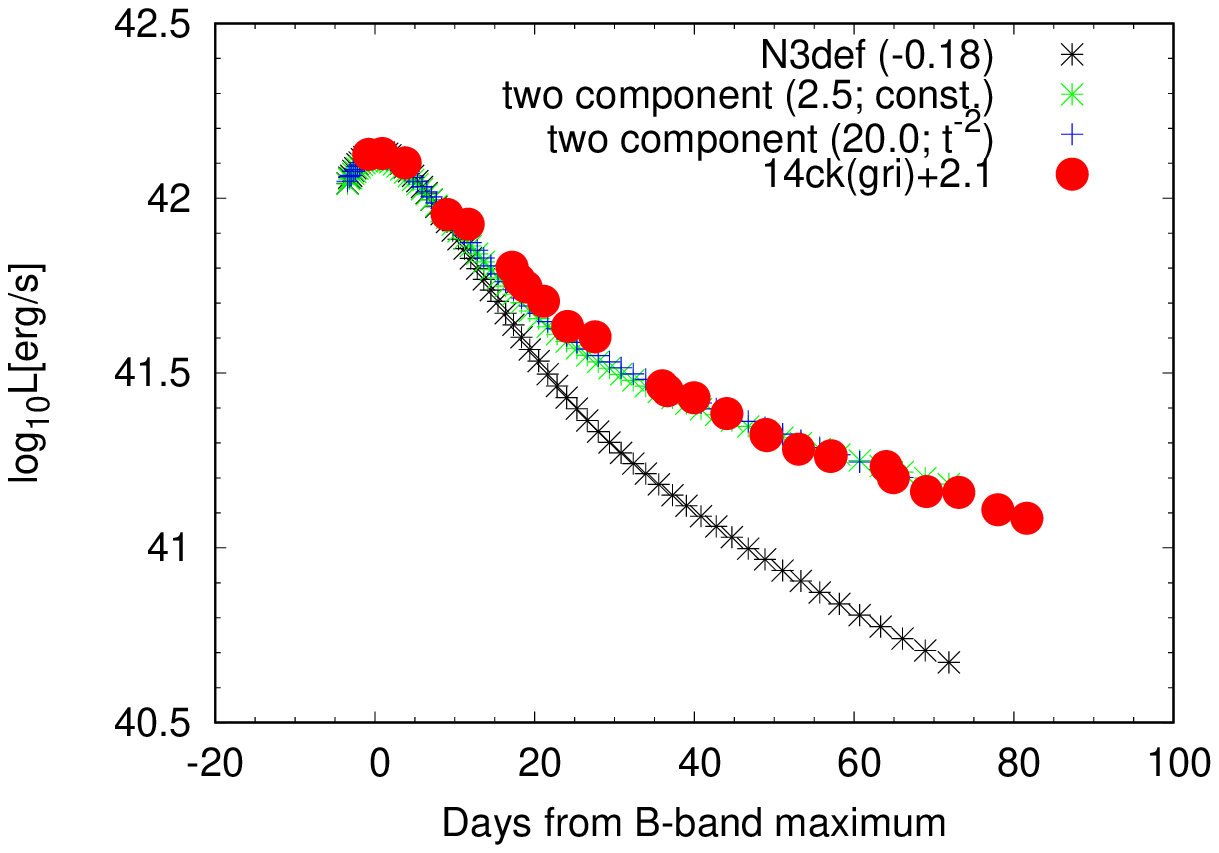}
 \end{minipage}
\end{center}
\caption{The same as Figure \ref{fig:km2_1} but for different SNe Iax.
}
\label{fig:km2_3}
\end{figure*}

\begin{figure*}
\begin{center}
 \begin{minipage}{0.44\hsize}
   \includegraphics[width=7.8cm]{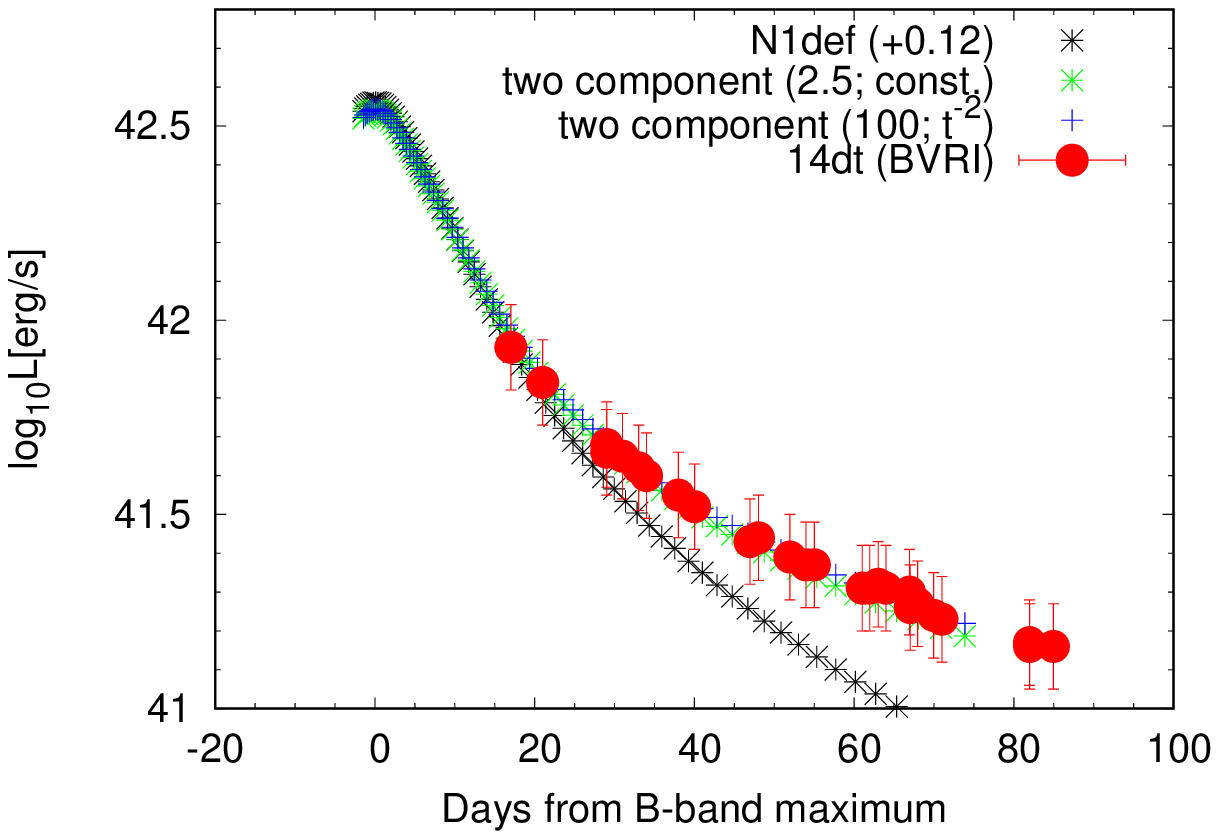}
 \end{minipage}
 \begin{minipage}{0.44\hsize}
   \includegraphics[width=7.8cm]{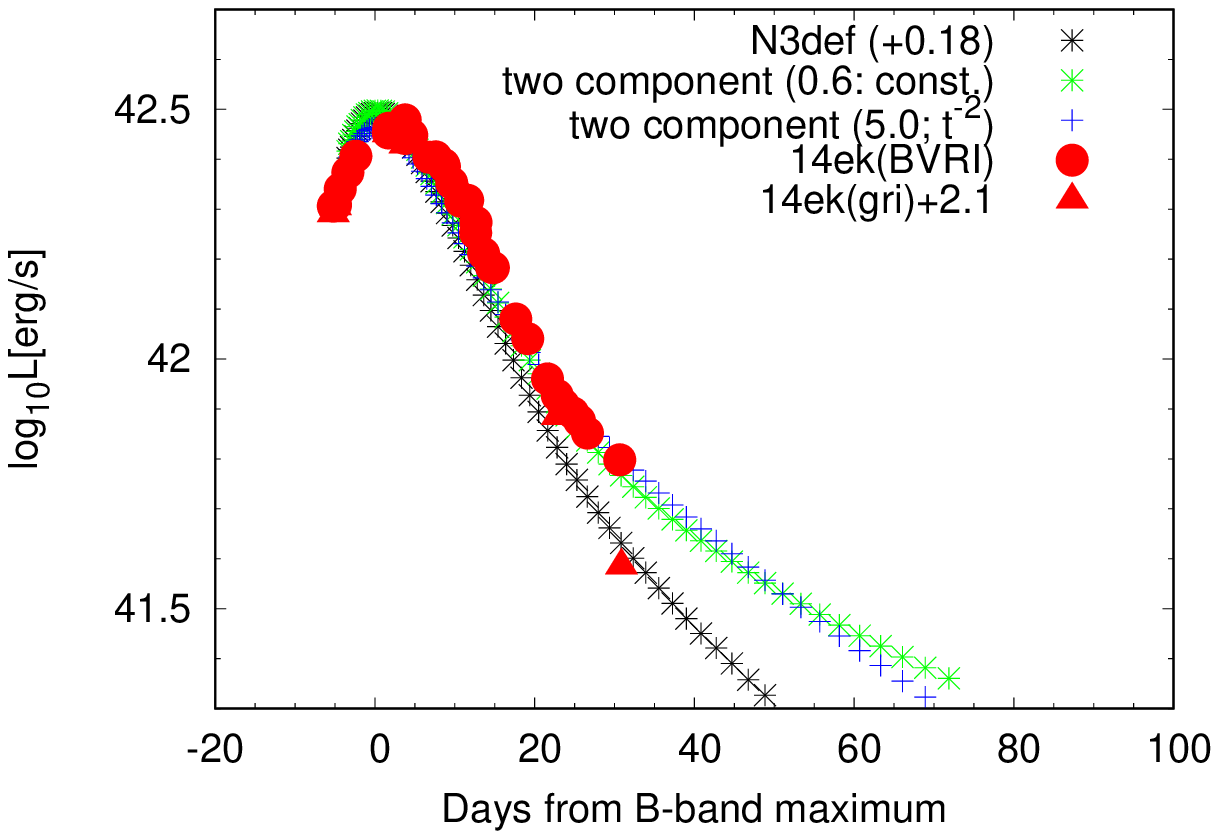}
 \end{minipage}
  \begin{minipage}{0.44\hsize}
   \includegraphics[width=7.8cm]{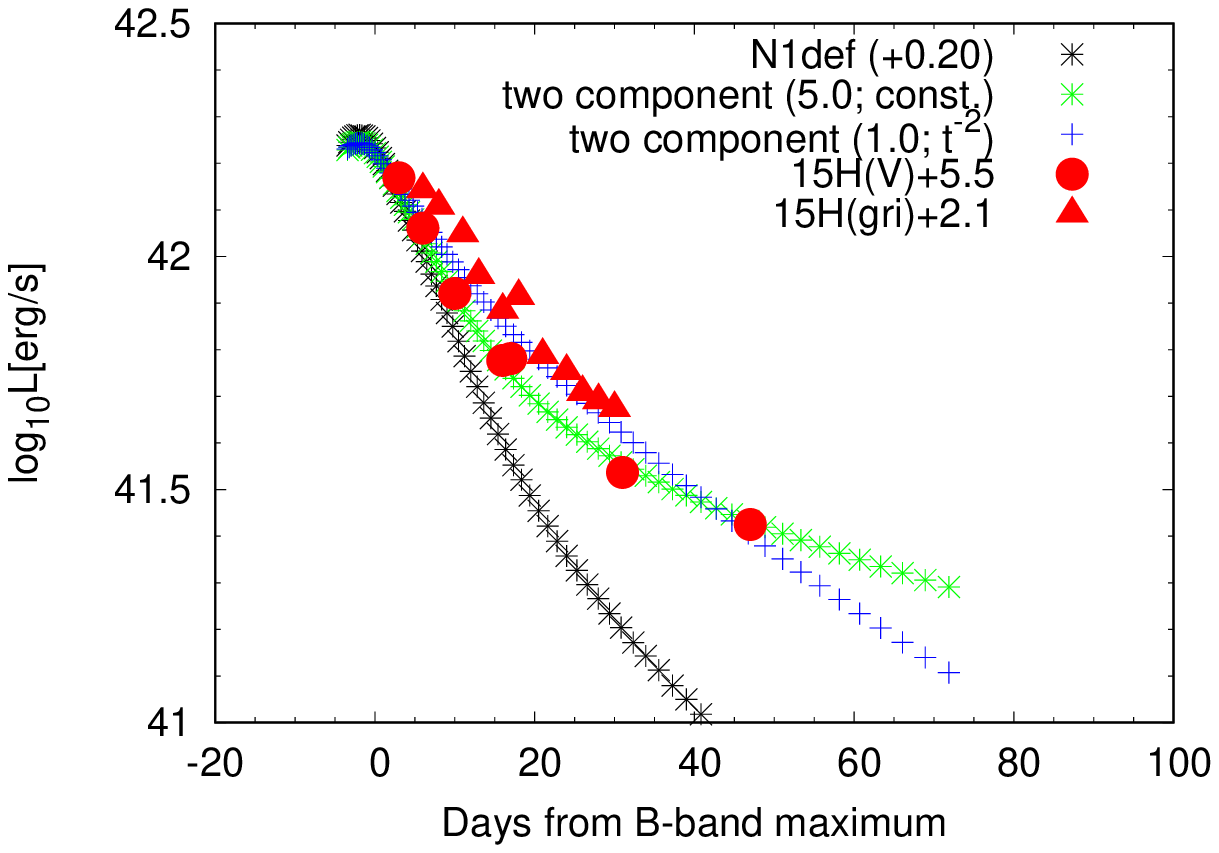}
 \end{minipage}
   \begin{minipage}{0.44\hsize}
   \includegraphics[width=7.8cm]{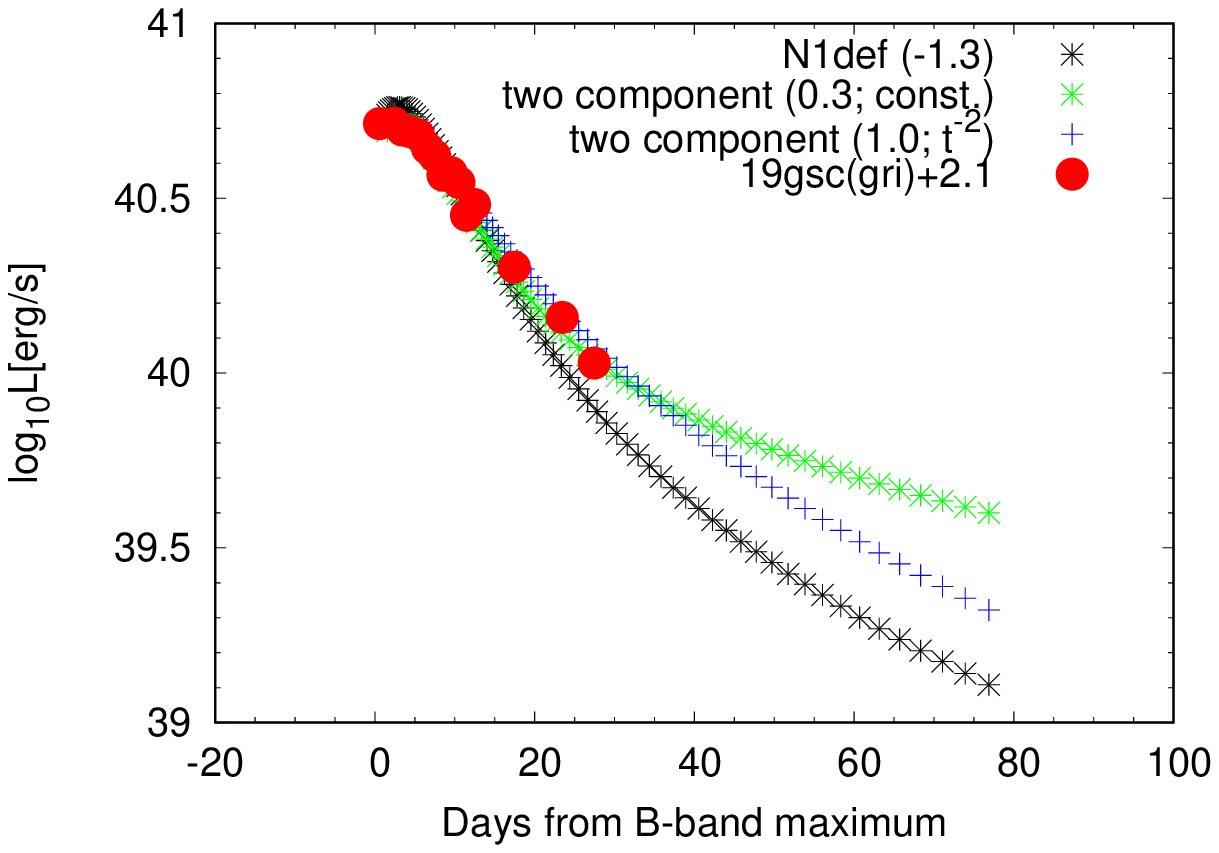}
 \end{minipage}
\end{center}
\caption{The same as Figure \ref{fig:km2_1} but for different SNe Iax.
}
\label{fig:km2_4}
\end{figure*}

\begin{figure}
\begin{center}
\includegraphics[width=12cm]{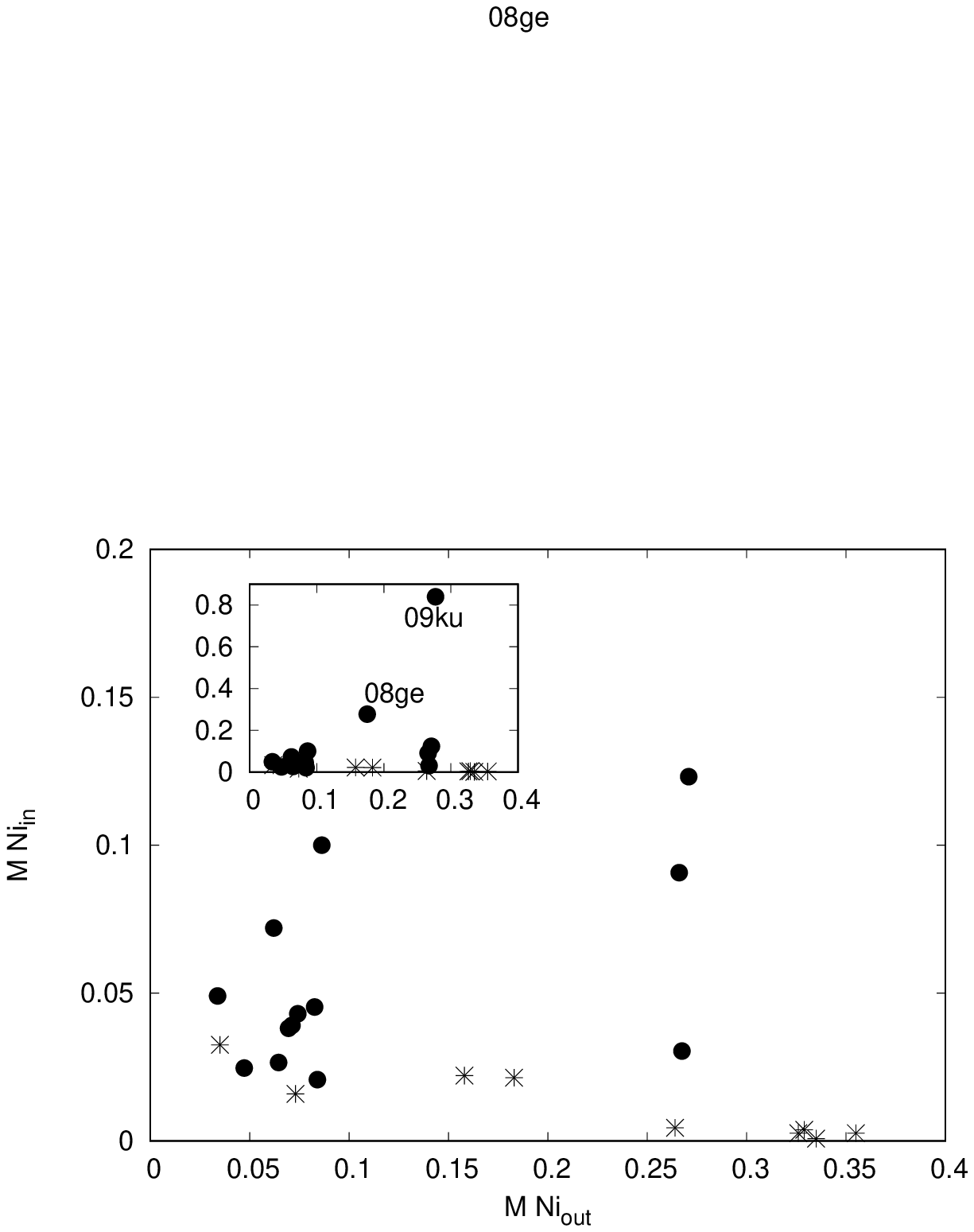}
\end{center}
\caption{Relationship between the $^{56}$Ni mass in the outer layer 
and that in the inner layer (filled circles).
The asterisk symbols denote the $^{56}$Ni masses calculated by \cite{Fink14}.
}
\label{fig:bol3}
\end{figure}

\begin{figure*}
\begin{center}
 \begin{minipage}{0.44\hsize}
   \includegraphics[width=7.8cm]{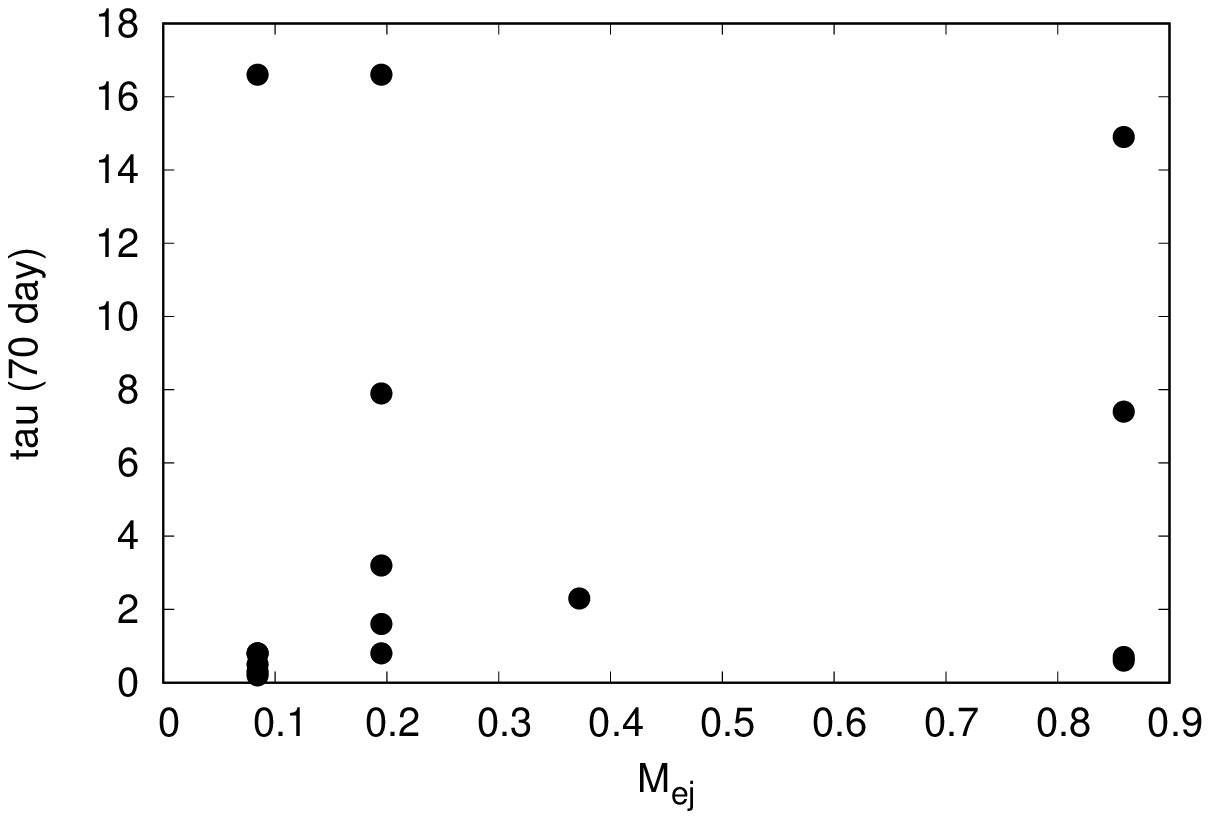}
 \end{minipage}
 \begin{minipage}{0.44\hsize}
   \includegraphics[width=7.8cm]{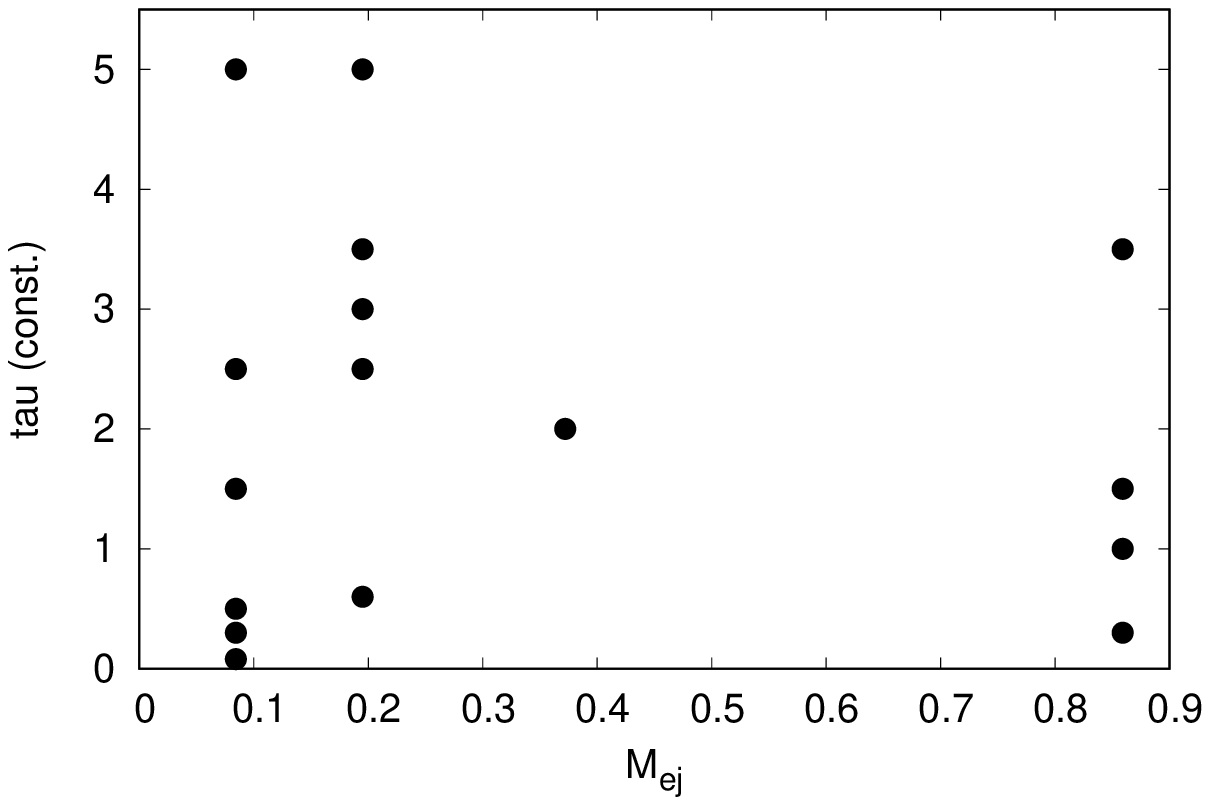}
 \end{minipage}
\end{center}
\caption{Relationship between the ejecta mass 
in the outer layer and the optical depth in the inner component. 
For the optical depth, the two cases are shown; $\tau = $ constant 
(right panel) or $\tau \propto t^{-2}$ (left panel). 
For the latter case, the optical depth is given at 70 days after 
the explosion.}
\label{fig:km2_5}
\end{figure*}

\begin{table}
  \tbl{Data sets of SNe Iax and the properties.}{%
  \begin{tabular}{cccccccccc}
      \hline
        SN &    L$_{peak}$     & $\tau$ &  [$(M_{\rm ej}/M_{\odot})^{2}/{E_{51}}$]$_{in}$   & M$^{56}$Ni$_{out}$ & M$^{56}$Ni$_{in}$ & Distance & E($B-V$) & Band & References \\
           & ($10^{42}$ erg/s) &          &          &                                      &  (Mpc)   &   (mag)  &      &            \\
      \hline
2002cx &  3.2     & 5.0 &  1.6  & 0.007 & 0.038 & 104.2 & 0.034  & $BVRI$ & 1  \\
2005cc &  1.1     & 3.0 &  0.8  & 0.084 & 0.020 & 37.0  & 0.0056 & $BVRI$ & 2,3,4,5 \\
2005hk &  3.6     & 3.5 &  7.9  & 0.065 & 0.029 & 49.2  & 0.11   & $BVRI$ $ugriz$ & 6,7,8 \\
2008A  &  5.4     & 3.5 &  7.4  & 0.267 & 0.030 & 70.0  & 0.0427 & $BVRI$ & 9  \\
2008ge &  2.5     & 2.0 &  2.3  & 0.175 & 0.278 & 16.8  & 0.0109 & $V$    & 10 \\
2008ha &  0.1     & 1.5 &  0.8  & 0.074 & 0.043 & 21.3  & 0.075  & $BVRI$ $ugri$ &  11,12 \\
2009ku &  5.0     & 0.3 &  0.6  & 0.278 & 0.829 & 370.0 & 0.0063 & $gri$  & 13 \\
2010ae &  0.3     & 0.1 &  0.3  & 0.062 & 0.072 & 13.1  & 0.62   & $BV$   $gri$  &  14 \\
2011ay &  6.0     & --  &  --   & 0.017 &  --   & 86.9  & 0.081  & $BVRI$ & 15 \\
2012Z  &  4.5     & 1.5 &  0.7  & 0.271 & 0.122 & 30.2  & 0.036  & $BVRI$ $ugri$ & 16,17 \\
2013en &  6.3     & 1.0 & 14.9  & 0.266 & 0.091 & 66.2  &  0.5   & $BVRI$ & 18 \\
2014ck &  3.2     & 2.5 &  3.2  & 0.083 & 0.045 & 24.4  & 0.48   & $gri$  & 19 \\
2014dt & 1.2--1.8 & 2.5 & 16.6  &  --   & 0.020 & 14.5  & 0.02   & $BVRI$ & 20 \\
2014ek &  3.2     & 0.6 &  0.8  & 0.071 & 0.039 & 99.5  & 0.054  & $BVRI$ $gri$  & 21 \\
2015H  &  --      & 5.0 &  0.5  & 0.038 & 0.049 & 60.57 & 0.048  & $V$ $gri$     & 22 \\
2019gsc & 0.025   & 0.3 &  0.2  & 0.086 & 0.100 & 39.8  & 0.01   &  $gri$ & 23 \\
2019muj & 0.5     & 5.0 &  16.6 & 0.047 & 0.018 & 32.46 & 0.023  & $BVRI$ & 24 \\
\hline
    \end{tabular}}\label{tb:datasets}
    \begin{tabnote}
(1) \citet{Li03}; (2) \citet{Ganeshalingam10}; (3) \citet{Silverman12};
(4) \citet{Lennarz12}; (5) \citet{Foley13}; (6) \citet{Sahu08};
(7) \citet{Holtzman08}; (8) \citet{Lennarz12}; (9) \citet{Hicken12};
(10) \citet{Foley10}; (11) \citet{Foley09}; (12) \citet{Stritzinger14}; 
(13) \citet{Narayan11}; (14) \citet{Stritzinger14}; (15) \citet{Szalai15}; 
(16) \citet{Stritzinger15}; (17) \citet{Yamanaka15}; (18) \citet{Liu15}; 
(19) \citet{Tomasella16}; (20) \citet{Kawabata18};
(21) \citet{Li18}; (22) \citet{Magee16}; (23) \citet{Srivastav20}; (24) This work.\\ 
\end{tabnote}
\end{table}

\section{Conclusions}\label{sec:conclusions}

We presented a long-term observation of SN 2019muj.
Based on the photometric properties of SN 2019muj, it belongs to the 
intermediate subclass between the bright and subluminous SNe Iax 
(see also \citealt{Barna21}). 
The spectroscopic features are similar to those of subluminous SNe Iax 
such as SNe 2008ha and 2010ae, with narrow absorption lines and slow 
expansion velocity. 
The similarity between the intermediate and subluminous subclasses in the 
spectroscopic features is in line with the previous work (e.g., 
\citealt{Stritzinger14}).
Our long-duration observations covering the epochs until $\sim$200 days 
after the B-band maximum show that SN 2019muj exhibits slow evolution of 
the light curve in the late phase. 
This behavior is similar to what was previously found for the bright 
SN Iax 2014dt.
Our observational data for SN 2019muj, including the multi-band data from 
the rising phase, serve as reference data set for a class of the 
intermediate SNe Iax, given that such well-sampled data are still limited 
for this class. 

We estimated that the $^{56}$Ni mass was 0.01 -- 0.03 $M_{\odot}$, the 
kinetic energy was $\sim$ (0.02 -- 0.19) $\times$ $10^{50}$ erg, and the 
ejecta mass was $\sim$ 0.16 -- 0.95 $M_{\odot}$. 
These values are consistent with the prediction of the weak deflagration 
model. 

The slow evolution in the late phase is best explained by an additional 
high-density component in the innermost layer. 
We reproduced the entire bolometric light curves of SN 2019muj and a 
sample of SNe Iax from the early to late phases, by adopting a 
phenomenological light curve model where we added an inner dense component 
powered by $^{56}$Ni, with the optical depth and $^{56}$Ni mass treated 
as free parameters. 
The model can explain the light curves of SNe Iax 
in general, which might support the existence of 
a high-density component in most, if not all, of SNe Iax.
The currently available data set is however not 
sufficient to further specify the origin of this inner component; this 
could be either the remnant WD itself, the slowly moving inner ejecta, 
or a wind launched from the WD. In any case, the remnant WD would likely 
create an inner dense component, and thus the present study supports the 
weak/failed deflagration model.

Our analysis also reveals possible shortcomings of applying 
the weak deflagration model as it is to the observations of SNe Iax; 
some important functions may still be missing in the present model.  
We find no clear correlation between the $^{56}$Ni masses in the outer 
ejecta and in the inner component, which is however predicted from the 
explosion simulations. 
Similarly, there is no relation found between the outer ejecta mass and 
the inner optical depth. 
In addition, the masses of $^{56}$Ni to explain the late-time luminosity are 
found to be generally larger than the model prediction. 
For the moment, it is not clear if the discrepancy could be remedied by 
considering detailed physical processes involved in the evolution of the 
remnant WD and its radiation. 
To constrain the nature of the bound WDs and test further the weak 
deflagration scenario, detailed observations of the candidate WDs with an 
extremely unusual atmospheric composition in the Milky Way (e.g., LP 40-365; 
\citealt{Vennes17}) would be useful.

\begin{ack}

We are grateful to the staff at the Seimei telescope and the Subaru 
telescope for their support.
The spectral data using the Seimei telescope were taken under the programs 
19B-N-CN02, 19B-N-CT01.
The late phase observation was performed with the Subaru telescope under S19B-055.
We are honored and grateful for the opportunity of observing 
the Universe from Maunakea, which has the cultural, historical and natural 
significance in Hawaii.
We thank the support staff at IAO and CREST who enabled the 2m HCT observations.
The IAO is operated by the Indian Institute of Astrophysics, Bangalore, India.
DKS and GCA acknowledge partial support through DST-JSPS grant 
DST/INT/JSPS/P-281/2018.
The authors also thank T. J. Moriya and M. Tanaka for insightful comments.
This research has made use of the NASA/IPAC Extragalactic Database (NED), 
which is operated by the Jet Propulsion Laboratory, California Institute 
of Technology, under contract with the National Aeronautics and Space 
Administration.
The spectral data of comparison SNe are downloaded from 
SUSPECT\footnote{http://www.nhn.ou.edu/\~{}suspect/} 
(\citealt{Richardson01}) and 
WISeREP\footnote{http://wiserep.weizmann.ac.il/} (\citealt{Yaron12}) 
databases.
This research has made use of data obtained from the High Energy 
Astrophysics Science Archive Research Center (HEASARC), a service of the 
Astrophysics Science Division at NASA/GSFC and of the Smithsonian 
Astrophysical Observatory's High Energy Astrophysics Division.
This work is supported by the Optical and Near-infrared Astronomy 
Inter-University Cooperation Program.
M.K. acknowledges support by JSPS KAKENHI Grant (JP19K23461, 21K13959).
K.M. acknowledges support by JSPS KAKENHI Grant (JP20H00174, JP20H04737, 
JP18H04585, JP18H05223, JP17H02864).
M.Y. is partly supported by JSPS KAKENHI Grant (JP17K14253).
U.B. acknowledges the support provided by the Turkish Scientific and 
Technical Research Council (T\"UB\.ITAK$-$2211C and 2214A).
\end{ack}

\bibliographystyle{apj}
\bibliography{iax}

\appendix 
\section{Bolometric Light Curves of A Sample of SNe Iax}\label{sec:app}

We assume that the $BVRI$ light curve of an SN Iax, as obtained by 
interpolating the SED and integrating the fluxes in the $BVRI$ bands, 
closely follows the bolometric light curve (see \S\ref{sec:bol}). 
However, the complete $BVRI$ data are not always available, and sometimes 
different filter systems are used to observe different SNe. 
We therefore apply the different bolometric correction to the data obtained 
with different combinations of the filter set. 

To obtain the bolometric correction for the data taken with different 
filter sets, we use the light curve data of well-observed SN Iax 2005hk, 
under the assumptions that the bolometric corrections are the same for 
all SNe Iax and do not evolve with time. 
First, we constructed the luminosity of SN 2005hk integrated with different 
sets of the photometric bands. 
The luminosity obtained is generally smaller than the $BVRI$ luminosity, 
because we only use the data in a limited number of the bands. 
Then, these light curves obtained for different filter sets are shifted 
vertically to match to the $BVRI$ light curve (which is assumed to be the 
bolometric light curve). 
The amount of the shift here is taken as the bolometric correction, and 
this is different for different filter sets. 

\begin{figure}
  \begin{center}
     \includegraphics[width=12cm]{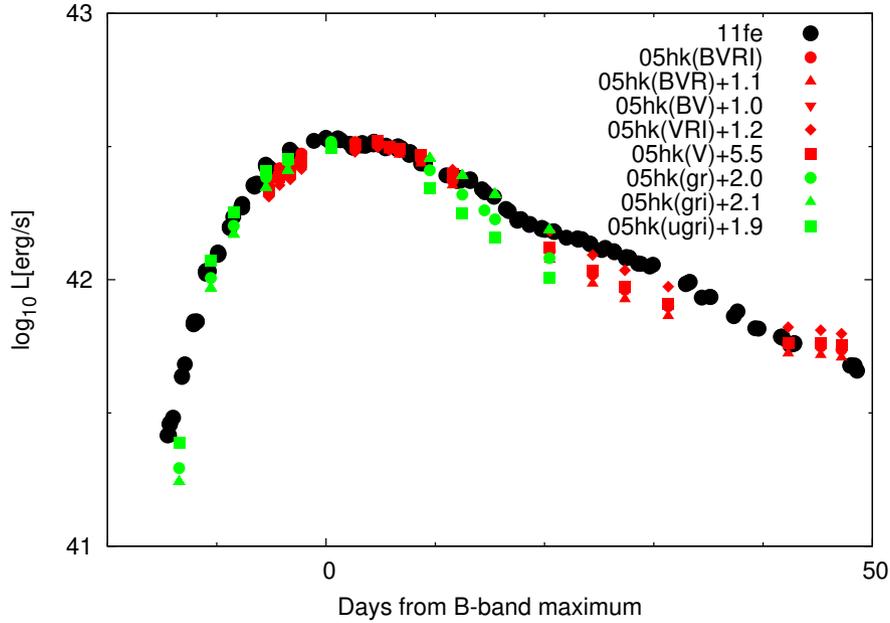}
       \end{center}
          \caption{The bolometric light curves of SN 2005hk constructed based on the data taken with different combinations 
          of the photometric bands.
	  The different symbol or color denotes the luminosity estimated using 
	  the different bands.
	  These light curves are shifted vertically with the amount indicated 
	  in the top-right portion of the panel to match with the $BVRI$ light 
	  curve around maximum light.
	  The $BVRI$ light curve is assumed to be bolometric, which matches to 
	  The bolometric light curve of SN 2011fe with an arbitrary shift in 
	  the vertical axis (black circles).
	  }\label{fig:05hk_fbol}
	  \end{figure}

\begin{figure}
  \begin{center}
     \includegraphics[width=16cm]{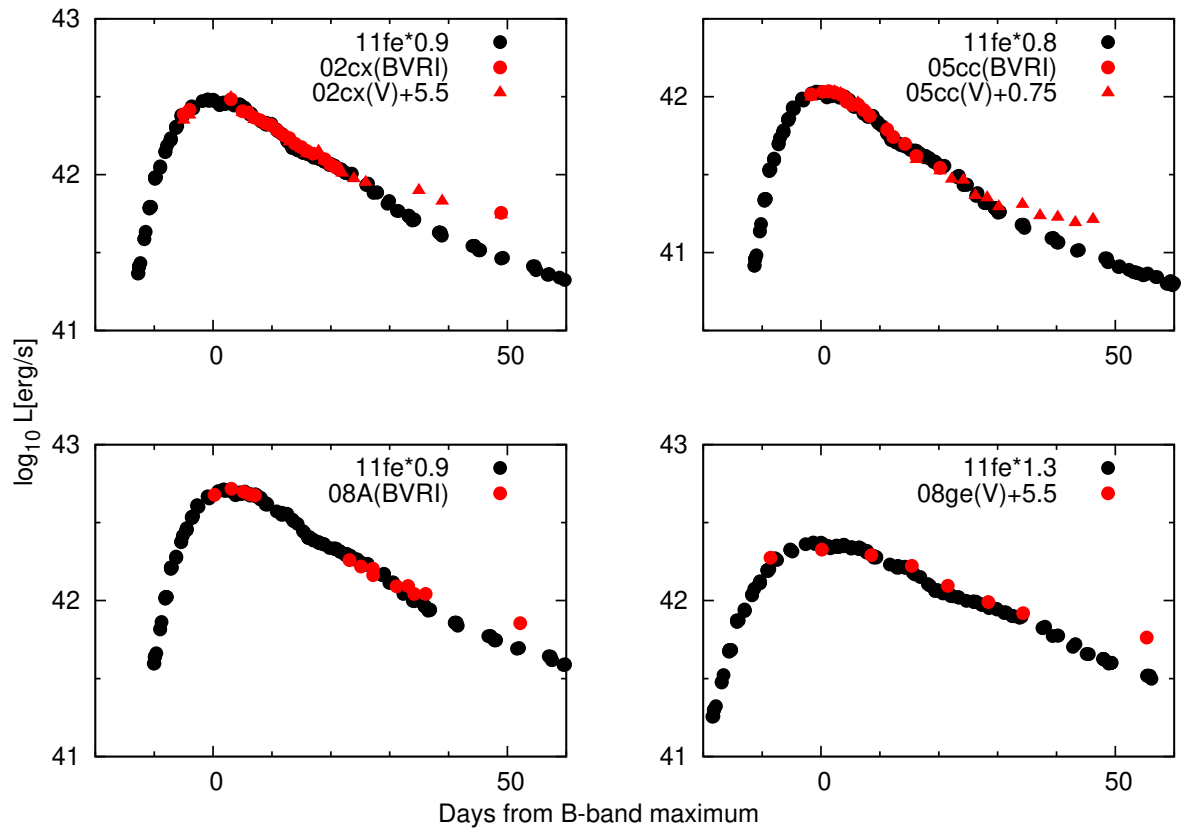}
       \end{center}
          \caption{The bolometric light curves of SNe 2002cx (left top),
	  2005cc (right top), 2008A (left bottom), 2008ge (right bottom).
	  We plot in the same manner as in Figure \ref{fig:05hk_fbol}.
	  }\label{test}
	  \end{figure}

\begin{figure}
  \begin{center}
     \includegraphics[width=16cm]{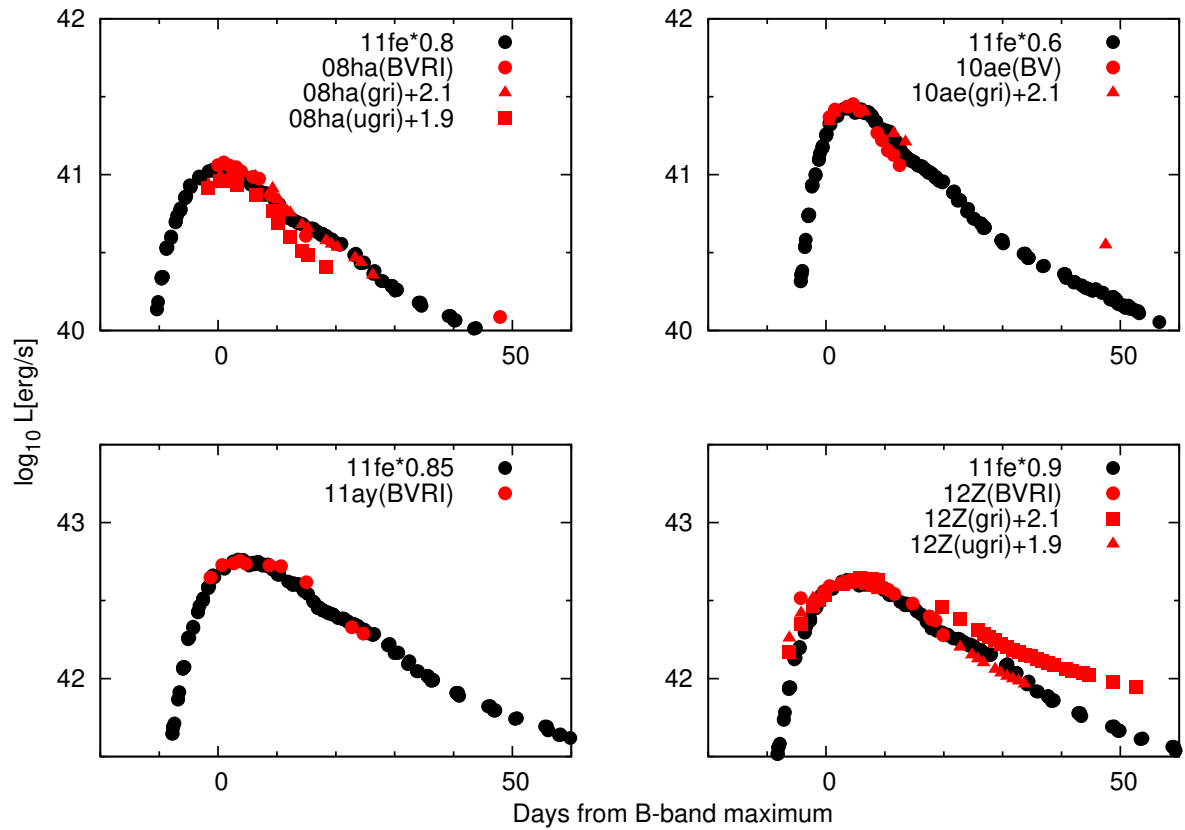}
       \end{center}
          \caption{The bolometric light curves of SNe 2008he (left top),
	  2010ae (right top), 2011ay (left bottom), 2012Z (right bottom).
	  We plot in the same manner as in Figure \ref{fig:05hk_fbol}.
	  }\label{test2}
	  \end{figure}

\begin{figure}
  \begin{center}
     \includegraphics[width=16cm]{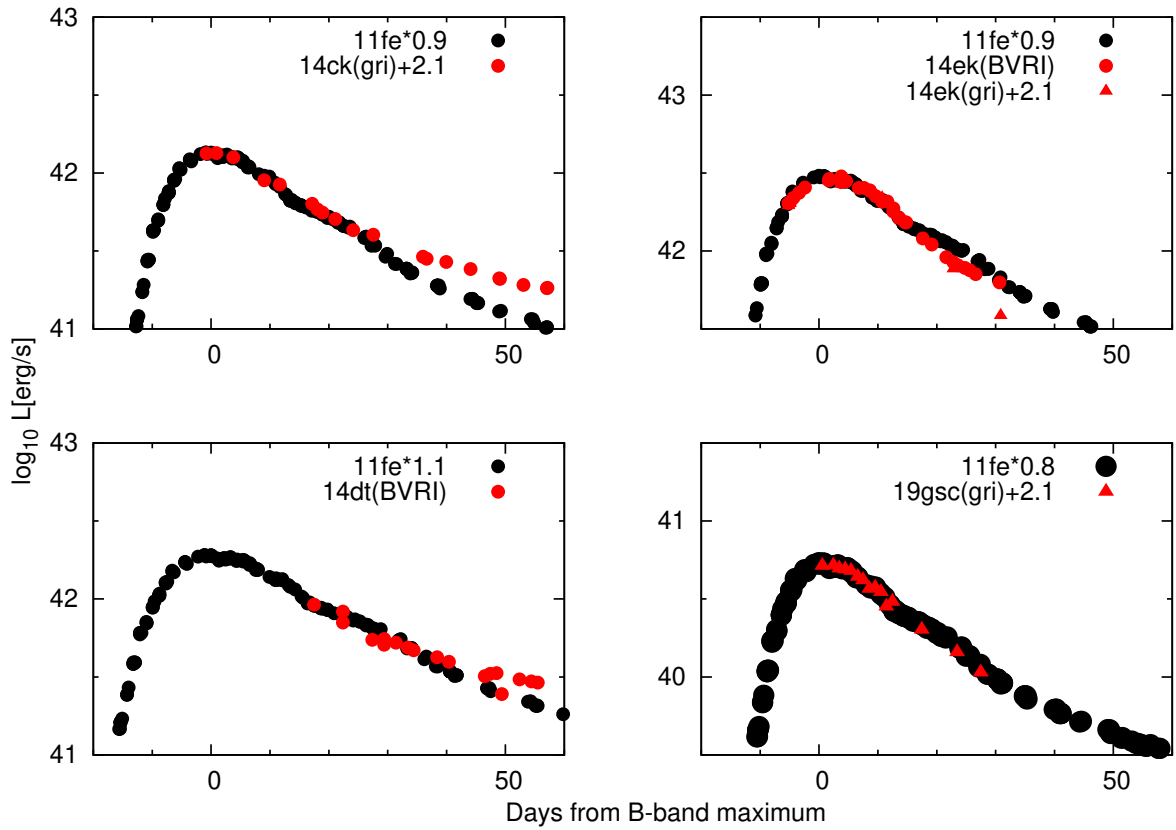}
       \end{center}
          \caption{The bolometric light curves of SNe 2014ck (left top),
	  2014ek (right top), 2014dt (left bottom).
	  We plot in the same manner as in Figure \ref{fig:05hk_fbol}.
	  The open circles of SN 2014dt is the bolometric light curve
	  assuming that the sum of fluxes in the $BVRI$-bands occupied
	  about 60 \% of the bolometric one.
	  }\label{test3}
	  \end{figure}

In Figure \ref{fig:05hk_fbol}, we compare the bolometric light curves of 
SN 2005hk estimated for different filter sets as described above. 
Also shown is the bolometric light curve of SN 2011fe as a cross-check. 
The bolometric corrections thus obtained for SN 2005hk are used for other 
SNe Iax, depending on the available filters for individual SNe Iax. 
The bolometric light curves of a sample of SNe Iax obtained in this manner 
are shown in  Figures \ref{test}, \ref{test2} and \ref{test3}.

\end{document}